\documentclass[aip,reprint]{revtex4-1}
\usepackage{amsmath}
\usepackage{graphicx}
\usepackage[utf8]{inputenc}
\usepackage[T1]{fontenc}
\usepackage{booktabs}
\usepackage{float}
\usepackage{filecontents}

\begin{document}


\title{The Feasibility of a Fully Miniaturized Magneto-Optical Trap for Portable
Ultracold Quantum Technology} 



\author{J.A. Rushton, M. Aldous and M.D. Himsworth}
\email[]{m.d.himsworth@soton.ac.uk}
\affiliation{School of Physics \& Astronomy, University of Southampton, Southampton, SO17 1BJ, UK}


\date{\today}

\begin{abstract}
Experiments using laser cooled atoms and ions show real promise for
practical applications in quantum-enhanced metrology, timing, navigation, and
sensing as well as exotic roles in quantum computing, networking and simulation.
The heart of many of these experiments has been translated to microfabricated
platforms known as atom chips whose construction readily lend themselves to integration with larger systems and future mass production. To truly make the jump from
laboratory demonstrations to practical, rugged devices, the complex surrounding
infrastructure (including vacuum systems, optics, and lasers) also needs to be
miniaturized and integrated. In this paper we explore the feasibility of
applying this approach to the Magneto-Optical Trap; incorporating the vacuum
system, atom source and optical geometry into a permanently sealed micro-litre
system capable of maintaining $10^{-10}$\,mbar for more than 1000 days of
operation with passive pumping alone. We demonstrate such an engineering
challenge is achievable using recent advances in semiconductor microfabrication 
techniques and materials.    
\end{abstract}


\maketitle 


\section{Ultracold Quantum Technology}
\label{intro}
Since the first demonstrations of atoms and ions at sub-millikelvin temperatures in the mid-1980s,
the field of atomic physics has been revolutionized by laser cooling and
trapping as it provides researchers with a method to probe some of the purest and
sensitive quantum systems available. This field is still highly productive and
recently has put significant emphasis on the practical applications of
this technology beyond the laboratory \cite{barrett2013mobile,kitching2011atomic}. 
It was evident very early on that
ultracold matter would be an indispensable tool in precise timing applications and a
recent demonstration \cite{Hinkley13092013} has shown extremely low
instabilities at the $10^{-18}$ level. The wavelike nature of atoms as they are cooled to
lower temperatures can be used to form atomic interferometers that
outperform optical counterparts in measurements of accelerated reference frames \cite{dickerson2013multi, canuel2006six, borde2002atomic, yver2003reaching}, which are important for inertial guidance systems, but can also provide sensitive measurements of mass,
charge and magnetic fields \cite{edwards2013ultracold,
budker2007optical,behbood2013real,wildermuth2006sensing}. Greater sensitivity
beyond the classical limit is possible via squeezed \cite{oblak2005quantum} and entangled states
\cite{napolitano2011interaction, pezze2009entanglement,appel2009mesoscopic},
which are also fundamental attributes for quantum computing
\cite{monroe2002quantum,kozhekin2000quantum}, and long distance quantum
networking \cite{duan2001long}. Ultracold matter has been used in the emerging
field of quantum simulation \cite{diehl2008quantum} and is an indispensable tool in determining
fundamental constants \cite{marion2003search}, testing general relativity
\cite{dimopoulos2007testing} and defining measurement standards \cite{borde2005base}. Many
researchers and industries believe such tools will be a major part of the
`second quantum revolution' in which the more 
`exotic' properties of quantum physics are applied for practical applications \cite{dowling2003quantum, UKquant}.

The field of ultracold matter has reached a maturity in both experimental methods
and theoretical understanding allowing experiments to begin leaving the
laboratory \cite{konemann2007freely, geiger2011detecting, van2007atomic}. These systems are bespoke, rarely
take up a volume less than a cubic metre and require a team of experts to operate. The many applications that
will benefit most
from ultracold quantum technology are likely to require far smaller and more
rugged devices which can be mass-produced and do not require the user to understand the internal operation in
detail. One can already see the opportunities made possible with the move to
microfabricated atom and ion traps \cite{groth2004atom,reichel2010atom,stick2005ion,brownnutt2006monolithic}, but
these firmly remain `chip-in-a-lab' \textit{components} rather than
`lab-in-a-chip' \textit{systems}. 

The miniaturization we envisage is analogous to that demonstrated by the recent
development of commercially available \footnote{Symmetricom SA.45 chip scale atomic clock.} chip-scale atomic clocks (CSACs), which have
shrunk a traditionally bulky optical spectroscopic system down to one smaller than a grain of rice
\cite{knappe2005atomic}. Some work has begun on miniaturizing the entire ultracold atom system, most noteably the 
backpack-sized iSense Gravimeter \cite{de2011isense}, but to achieve the CSAC level of sophistication, size and robustness
in ultracold technology will require at least another decade of development.

The trapping and cooling of hot vapour-phase atoms or ions below
millikelvin temperatures is the first stage in all ultracold experiments,
therefore the miniaturization of the system known as the Magneto-Optical Trap \cite{chu1998manipulation}
(MOT) would be a significant step forward towards our goal. Several academic and commercial
research groups have begun looking at the various ways the MOT can be
miniaturized using machined glass chambers \cite{salit2012progress}, conical
retro-reflectors \cite{xu2008realization, shah2011miniature}, and etched
multi-section silicon and glass substrates \cite{salim2011compact}. Most of
these demonstrations are small-scale versions of standard MOTs, with only the
last device beginning to redesign the system from a microfabricated and
integrated approach. 

In this study we explore the feasibility of miniaturizing
and integrating the ultra-high vacuum system, atom source and MOT optics into
a centimetre-scale device. This will be achieved by using recent advances in
materials and techniques adapted from the semiconductor and MEMS industries 
used in wafer-level mass production. We will refer to the 
device as a `MicroMOT' because the internal volume is on the scale 
of micro-litres compared to the typically litre-sized standard MOTs.
The initial target operational lifetime is set at 1000 days, as this would be 
at the lower end of a typical commercial service life whilst still presenting a significant 
challenge. We also aim to maintain an internal vacuum of $10^{-10}$\,mbar 
under normal atmospheric external conditions, and do so with only passive pumping elements and thus no power. 
Our objective is to focus on this as an engineering challenge from which a mass-producible 
technology can be developed, thus avoiding bespoke systems which may only be suitable for
proof-of-concept purposes. 

In Section \ref{MOTsec} we describe a typical Magneto-Optical Trap system,
its construction, and how it can be miniaturized. In Section \ref{atomsec} we
discuss the source of vapour phase atoms and how to control them. In Section
\ref{UHVsec} we explore solutions to provide pumping, prevent
permeation, limit leaks, and overcome outgassing. In Section \ref{protosec}
we bring the above technologies together to design a prototype Micro-MOT. In
Section \ref{dissec} we discuss the assumptions made in the study and highlight areas for further research.

\section{The Magneto Optical Trap System}
\label{MOTsec}

Nearly all cold atom experiments begin with a Magneto Optical Trap of
which a typical design comprises an Ultra-High Vacuum (UHV, <10$^{-9}$ mbar)
chamber with internal volumes of around a litre with numerous optical ports,
atom sources, gauges and pumps attached. UHV is obtained by thorough cleaning of
the polished glass and metal (typically stainless steel) components. The entire system is assembled and evacuated using
roughing and turbomolecular pumps down to around 10$^{-7}$ mbar. It is then baked 
in the vicinity of 200$^\circ$C for several days whilst being evacuated by 
ion and sublimation pumps and, once cooled, will obtain vacua in
the region of 10$^{-10}$ mbar. Obtaining vacua much beyond this, in the
extreme high vacuum (XHV) regime, can be very difficult and may require getters, 
cryogenic pumps, deeper cleaning regimes and alternative chamber materials.

\begin{figure}
\includegraphics[width=0.45\textwidth]{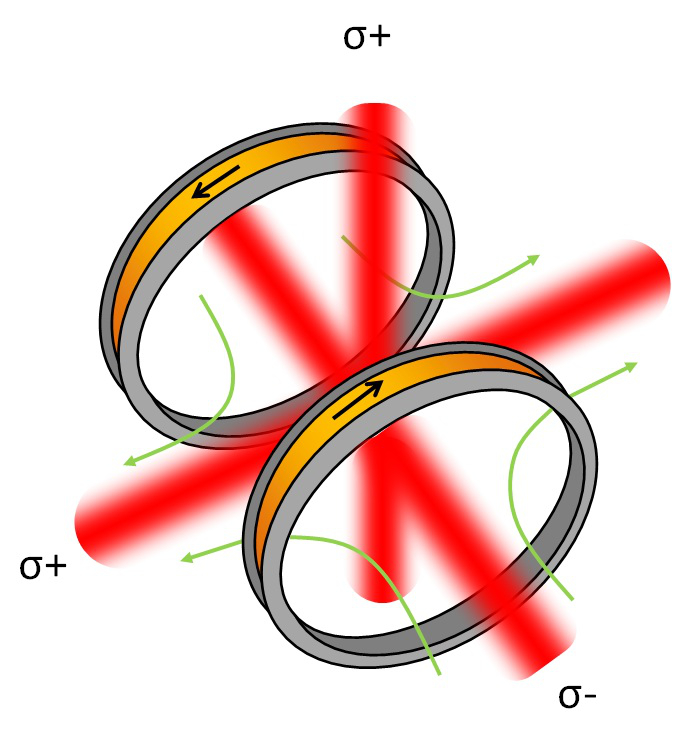}
\caption{\label{stdMOT} The standard MOT geometry. The laser
polarizations are indicated in text and the magnetic field direction in green
arrows.}
\end{figure}

Once UHV is obtained, the MOT is 
formed of several stabilized and finely-tuned laser beams that are retro-reflected along each
Cartesian axis intersecting at 
the zero of a quadrupole magnetic field (see Figure \ref{stdMOT}). Vapour-phase
atoms are released into the chamber, cooled, trapped, and finally manipulated
for their intended task. Typically $10^{7}$ atoms are trapped in a dense cloud
with diameters usually below 1\,mm and, for the majority of experiments
(excluding long freefall experiments), the atoms rarely move more than a few millimetres
away from this point. The past decade has seen the emergence of atom chips which
allows for manipulation of atoms microns away from surfaces using high magnetic
field gradients, created by microfabricated wires \cite{folman2000controlling}. 

This raises the question to why
such a large vacuum system is required? The answer is that without resorting to bespoke designs the pumps and gauges
one can purchase for UHV systems are very large, and regardless, using current approaches the system is still difficult to reduce
below the size of a shoebox.  Typically these are far too bulky,
expensive, and labour intensive to mass-produce and so an alternative 
architecture and manufacturing approach is required, starting with the MOT
geometry.
\vspace{3mm}

\begin{figure}
\includegraphics[width=0.45\textwidth]{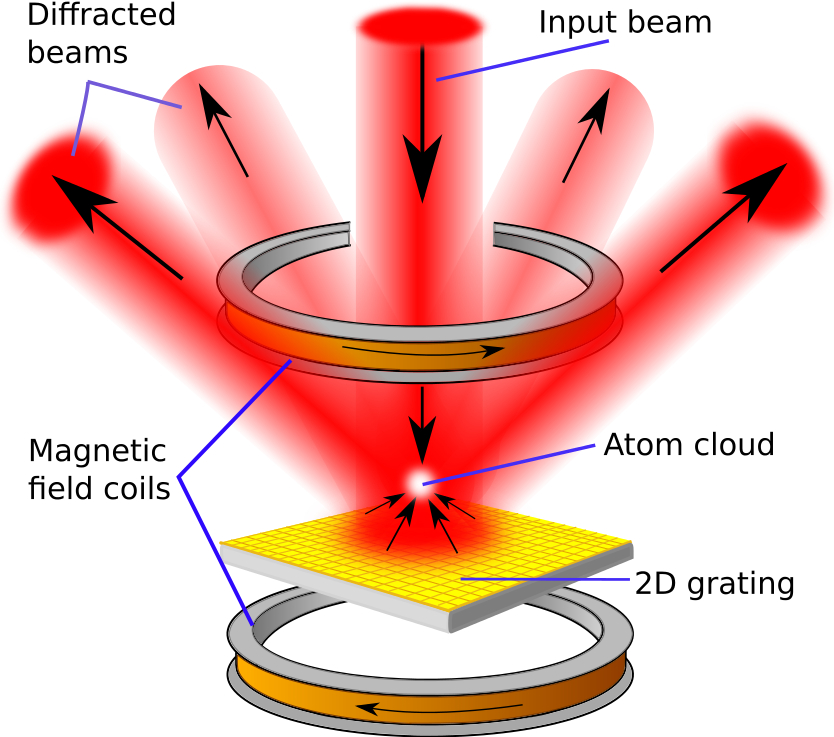}
\caption{\label{GMOT} A grating MOT geometry.}
\end{figure}

For an integrated device the `standard' geometry presented above is impractical
due to the need for many optical ports, complex alignment, large volumes,
numerous fragile optical elements and the difficulty in bringing the atoms close
to an atom chip surface. Several alternative geometries have been proposed
including the mirror-MOT \cite{wildermuth2004optimized}, pyramid-MOT
\cite{lee1996single,arlt1998pyramidal}, and tetrahedral MOT
\cite{vangeleyn2009single}. The latter two are attractive as they need only a
single incident beam, and both are suitable for
microfabrication. Miniaturized pyramid MOTs, however, suffer from low atom
capture rates due to the small volume in which the beams overlap
\cite{trupke2006pyramidal, pollock2009integrated},
significant backscatter making the atoms difficult to detect \cite{pollock2011characteristics}, 
and the geometry making transfer of the atoms to magnetic surface traps
non-trivial. A recently demonstrated version of the tetrahedral-MOT using a
planar grating as a reflector (which we refer to as the `G-MOT', see Figure \ref{GMOT}) can capture a
large number of atoms, has lower backscatter \footnote{This can be mitigated with background free imaging \cite{ohadi2009magneto}}, 
and can be easily integrated with
atom chip structures \cite{nshii2013surface}. Some disadvantages include the effect
of the grating on the wavefronts and polarizations of the manipulation beams \cite{lee2013sub}, and added difficulty in situations
which require several widely-spaced wavelengths. Nevertheless, the G-MOT appears
to be the most suitable geometry for microfabricated devices. 
\vspace{3mm}

For most cold atom experiments, the practical starting number of trapped atoms is
on the order of $ N_{t}=10^{6}$, with a lower limit \cite{hoth2013atom} around
$10^{4}$. The G-MOT characteristics show that the beam overlap volume,
$V_{t}$ (cm$^3$), follows the scaling law of $ N_{t}=4\times10^{7}V_{t}^{1.2}$,
resulting in a minimum practical volume of  $0.045$\,cm$^{3}$. This is equal to a
0.65\,cm diameter, uniformly illuminated, beam\footnote{Several studies have shown
the scaling laws for various beam overlap volumes, however we highlight that
none have considered the effect of unsymmetrical volumes for which reducing one
dimension could be beneficial in wafer-level microfabrication by reducing
chamber size whilst maintaining a large capture volume.} forming a pyramidal volume. To ensure the correct
number of atoms and to take into account the effects of non-uniformly shaped beams, a
pragmatic beam diameter would be 1\,cm. Experiments requiring degenerate gases may 
require at least $10^{8}$ atoms to ensure a stable phase density for condensation and also improve detection. 
This would warrant much larger capture volumes than described here. 

We assume that the device would incorporate an atom chip structure so that the
atoms are trapped and manipulated with magnetic fields close to the surface, therefore
the dimensions of the device have little impact on the measurement. Studies have shown that laser cooling
close to surfaces begins to show losses as the atom-surface separation decreases below 1\,mm \cite{pollock2011characteristics, huet2011magneto}. Thus, assuming a typical MOT cloud with a diameter less than 1\,mm, a lower limit on the vertical dimension would be 3\,mm.
We aim to explore this issue in another study, but point out here that for a
1\,cm diameter GMOT beam, $\sim94$\% of the overlap volume is contained within the
first 3\,mm from the grating surface. In other matterwave experiments which manipulate the atoms during free fall, the interaction time may
be limited to several milliseconds in a 3\,mm thick chamber. For example, if we assume 
our trapped species is rubidium which has been cooled to the Doppler limit of 146\,$\mu$K 
(r.m.s. speed of 20\,cm\,s$^{-1}$), and the atoms have 1\,mm to travel before losses 
occur, then the maximum interaction time is 5\,ms neglecting gravity. Sub-Doppler cooling can increase this by a
factor of 3 to 4,  which may be acceptable in many situations. Optical lattices can increase the interaction time of matterwave
interferometers without drastically increasing the dimensions of the
system \cite{kovachy2010optical}. These guided matter waves are a promising technique which could 
 aid miniaturization and improve sensitivity.

\section{Atom Source and Control}
\label{atomsec}
The atomic species to be cooled and trapped can be sourced either from a hot
vapour, or captured from an atomic beam. The latter is usually produced 
from a hot Knudsen oven, and requires additional cooling to obtain a suitable capture
efficiency in the MOT, usually via a Zeeman slower \cite{phillips1982laser} or chirped cooling \cite{phillips1985laser}.
Our chip based system will be far
too small for such slowers as they require tens of centimetres for adequate deceleration,
although they can be made smaller using bichromatic force techniques \cite{liebisch2012atom}. Loading from
a background vapour is a common method which results in reasonably fast loading 
rates, but requires a vapour pressure greater than UHV, resulting in increased 
 collisions and decoherence during subsequent
manipulation \cite{arpornthip2012vacuum}. Therefore the vapour pressure must be controllable on short times scales, ideally within a second.
A common method to achieve this uses a MOT cooled in two dimensions (2D-MOT) in one chamber separated from
a 3D-MOT in another via a narrow conductance channel \cite{berthoud1998continuous}. The 2D-MOT
chamber may be kept at a high vapour pressure so that it may load many atoms into a low velocity beam directed into the
higher vacuum 3D-MOT chamber. This technique has been used in the \textit{ColdQuanta} miniaturized
BEC system \cite{salim2011compact}. 
Such multichamber systems are likely to be necessary for obtaining BECs which require a higher level of vacuum, however
in this study we aim solely to produce a cold non-degenerate cloud of atoms, 
concentrating on loading a single MOT from a room temperature vapour, and to
control it on short timescales. 
\vspace{3mm}

Each laser cooled species has different chemical properties which bring
different challenges. In this study we look at rubidium as it is ubiquitous
across the whole scope of cold atom experiments, and poses the challenge of a
vapour pressure which is too high at room temperature for efficient trapping. Species with
lower vapour pressure, such as strontium, could be easier to use as they do not
endanger the vacuum, but the high temperatures needed to obtain a suitable background pressure results in less efficient trap loading, and so may
require the additional cooling mechanisms discussed above. Rubidium \cite{steck2001rubidium} melts at 39$^\circ$C and at room temperature has a vapour pressure of
$5\times10^{-7}$\,mbar.  This results in significant
collisional rates with trapped atoms and also excessive fluorescence, making the
detection of the cold atoms very difficult. Moreover, the very small volumes
inside the chips, and the lack of active pumping, quickly results in vapour
saturation. A sufficient vapour pressure to load a rubidium MOT is $\sim 10^{-8}$\,mbar, but one must reduce this by an order of magnitude for any decoherence-sensitive
measurements. Therefore, one must have a method to carefully regulate
the flow of rubidium into the MOT chamber. 

The past decade's development of CSACs has provided a range of methods to introduce alkali atoms into microfabricated devices. These sources include pure metal \cite{liew2004microfabricated,vecchio2010dispensing}, 
alkali compounds \cite{liew2004microfabricated}, wax pellets \cite{radhakrishnan2005alkali}, alkali azides \cite{liew2007wafer}, and alkali-enriched glass \cite{gong2006electrolytic}. Most are not suitable for UHV
or result in poorly controlled, or limited lifetime, sources. Pure rubidium is
not suitable unless it is sealed away during fabrication as its high pressure vapour will
ruin vacuum at the elevated temperatures required for baking and bonding. Commercial alkali dispensers, such as \textit{SAES Getters Alkali Metal Dispensers} (AMDs) and \textit{Alvatec Alvasources} are alkali compounds which are stable up to temperatures of
300-600$^{\circ}$C. AMDs are chromates combined with a Zr-Al getter
material held in a nichrome dispenser \cite{bartalini2005full}. Heating of the AMDs
results in a reduction reaction releasing pure rubidium and some additional
gases which are gettered away. \textit{Alvasources} are alkalis alloyed
with `poor' metals, such as bismuth, which form stable compounds with higher
sublimation temperatures than their constituent elements. They also result in
far less residual gas than AMDs \footnote{www.alvatec.com}, albeit at a higher cost. Both of these sources can be controlled
with Joule heating, but they can also be activated with a focused laser \cite{griffin2005fast, douahi2007vapour}, removing 
the need for electrical feedthroughs and reducing
the heat transfer to the chip \footnote{Integrating high power focused beams
into a chip scale system would not be trivial and the beam must be moved once
sections of the dispenser are depleted. An alternative heating method could include induction heating.}. 

Rubidium vapour will reach saturation very quickly within micro-litre volumes,
especially as the previously mentioned sources may be difficult to
control accurately, so a system to pump away the vapour must be incorporated.  
Glass and metals are effective pumps for alkali atoms: surface studies have found binding
energies around 3\,eV and extremely high pumping rates \cite{arpornthip2012vacuum} of 
10$^{3}$\,l\,s$^{-1}$cm$^{-2}$. Studies looking at vapour cell
coatings \cite{stephens1994study} have highlighted a significant `curing time' after filling, during which the vapour pressure stabilizes due to strong chemisorption \footnote{In that study they found that Pyrex took an order of magnitude
longer to cure than sapphire. The glass used in this study is likely to have
properties in between these materials.}. After the surface is saturated 
the adsorption energy drops to $\sim0.5$\,eV and is thus only weakly physisorbed. If we assume the MicroMOT produces a 10\,second pulse of rubidium
every minute, with a peak pressure of $10^{-8}$mbar, which is
pumped away at 1\,l\,s$^{-1}$, one would require a total of $10^{19}$ atoms (about 1\,mg) to last for our 1000 day target. 
A typical monolayer is around $5\times10^{14}$\,cm$^{-2}$, so one cannot rely on surface 
pumping alone if it cannot be degassed regularly \footnote{The effect of hard optical coatings on
the glass, such as MgF$_2$ to reduce reflections or ITO for heating, may also reduce
the effectiveness of glass as a pump.}. We note that the limited surface area can be 
increased with materials such as
aerogel, porous silicon, zeolites, and anodic alumina.

An obvious and effective method to control the vapour is by simply reducing the temperature of the 
MicroMOT. To get to $10^{-10}$\,mbar one must cool rubidium to $-30^\circ$C. This can be
accomplished by cooling the entire chip or with an integrated `cold finger',
such as a micro-peltier device \cite{bottner2002thermoelectric}. This latter method will avoid rubidium condensation on 
critical features such as the windows or reflectors, and also avoid water accumulating on external surfaces.
The pumping of alkali metals by getter films
has been reported to be negligible \cite{scherer2012characterization, Porta1968alkali}, but little data is available \footnote{Scherer et a l\cite{scherer2012characterization} state that NEGs do not pump rubidium at all, 
whereas della Porta et al \cite{Porta1968alkali}
state that alkali atoms do not affect the pumping speed for other gases. We
suspect there is some pumping due to the high reactivity of alkali atoms with
metals and with most adsorbed gases.}, so may not be useful in its regulation.
Many atom chips require gold films for reflective surfaces and conductors and it is known in the field that these may
degrade over time when exposed to a hot rubidium source. The
phase diagram \cite{pelton1986rb} between gold and rubidium shows a stable alloy forms around 500$^\circ$C.
Therefore one can use a heated gold surface to
pump away excess rubidium. Another method could utilize the rubidium/bismuth
alloying effect mentioned earlier as a thermally controlled pump, but one must
be wary of the low melting point of this metal (271$^\circ$C) during fabrication. Both alloying methods work for all alkali metals but, as shown in Table \ref{alloy}, these occur at different temperatures.

A common method to quickly control the vapour pressure whilst remaining at room temperature
is Light Induced Atomic Desorption (LIAD) \cite{torralbo2013light}. This technique involves the
illumination of metal or glass surfaces with non-resonant ultraviolet light (UV) in 
order to increase the desorption rate of physisorbed alkali atoms. 
The exact mechanism by which this occurs is still under debate \cite{hatakeyama2005classification,rkebilas2009reexamination,villalba2010reply}.
Once the UV light is extinguished the desorption rate reduces so that atoms can return to the surfaces. 
This reloading of the atom sources means that the total number of atoms in the device can be reduced through recycling.
Studies have shown an order of magnitude improvement of MOT loading rates with this 
 technique \cite{atutov2003fast,anderson2001loading},
and it has been used to make BECs, which are very sensitive to background gas collisions,
in a single chamber \cite{xiong2013production}. In chip-scale systems
the surface area is far too small for effective use of LIAD \cite{du2004atom}
but, as mentioned earlier, one can introduce high surface area 
materials \cite{burchianti2004light,villalba2010light} providing they can be
degassed sufficiently prior to encapsulation. 

\begin{table}
\caption{\label{alloy}Gold and bismuth alkali alloys with
1:1 compositions for use as alkali pumping mechanisms. Many of the phase diagrams
 exhibit several phases with additional alloys forming above and below these
temperatures and the reader should refer to the original sources. The
approximate values are due to indistinct alloying temperatures. }
\begin{tabular}{lccc}
\toprule
Alkali &\hspace{5mm} Au-M alloy&\hspace{5mm} Bi-M alloy\\
metal (M) &$^\circ$C&$^\circ$C \\
\midrule
Li \cite{pelton1986li,sangster1991bili}& $\sim$660 & $\sim$400\\
Na \cite{pelton1986gold,sangster1991bina}& 372 & 444\\
K \cite{pelton1986k,petric1991bi}& 532 & 355 \\
Rb \cite{pelton1986rb,pelton1993bi}& 498 & 376 \\
Cs \cite{pelton1986cs,sangster1991bics}& 585 & 390 \\
\bottomrule
\end{tabular}
\end{table}

For any pumping mechanism the production of rubidium from the source should be
well controlled to ensure consistent loading of the MOT and to prevent permanent vapour
saturation. If the source reactively produces hot vapour at unpredictable rates,
due to material or heating inhomogeneities, then additional mechanisms are needed
to control the flow. Separating two chambers of different pressures is a
common challenge in UHV systems, as discussed earlier in
2D/3D MOT loading, and can be achieved by carefully limiting the
gas conductance between them with a narrow channel. 
A channel 1\,mm long with a cross-section of $100\times100$\,$\mu$m, can maintain UHV in the
MicroMOT chamber at room temperature \cite{o2005user} whilst the source chamber is at saturation pressure, as
long as there is a pumping rate greater than 0.1\,l\,s$^{-1}$ in the laser cooling chamber. Locally heating the source chamber by 100$^\circ$C will 
sufficiently increase the vapour pressure for loading the MOT. The narrow aperture also
leads to a `beaming effect' which may aid the loading of the trap.

\section{UHV in a chip}
\label{UHVsec}

Table \ref{uhvprop} highlights the various challenges in terms of leak,
permeation, and outgassing rates that must be tackled to realize sealed passive
UHV chips, and compares them to those required by typical UHV systems. Reaching many of these values, especially those for noble gases, may seem
unachievable, however we have identified methods to do so by careful choice
of materials, fabrication processes and also structural features. 

\begin{table}
\caption{\label{uhvprop} General characteristics of standard UHV MOT
systems, and those for the MicroMOT}
\begin{tabular*}{0.45\textwidth}{@{\extracolsep{\fill} }lcc}
\toprule
& Standard&  MicroMOT\\ 
\midrule
Internal volume (l) & >1 & <$10^{-3}$\\
Lifetime (days) & indefinite& 1000 (target)\\
Pump rate (l\,s$^{-1}$) & >20& <1 \\
Leak rate (mbar\,l\,s$^{-1}$) & <$10^{-11}$  & \parbox[t]{2cm}{<$10^{-19}$(Ar)\\<$10^{-14}$(N$_2$)}\\
\parbox[t]{3cm}{Outgassing rate\\(mbar\,l\,s$^{-1}$cm$^{-2}$)} & <$10^{-11}$ & \parbox[t]{2cm}{<$10^{-21}$(He)\\ <$10^{-16}$(H$_2$)}\\
Permeation rate\footnote{For helium using Equations \ref{conperm} and \ref{vaceq}.} (cm$^{2}$s$^{-1}$) & <$10^{-7}$ &  <$10^{-17}$\\ 
\bottomrule
\end{tabular*}
\end{table}

Vacuum encapsulation of microfabricated devices is a large and mature industry and
nearly all MEMS devices require some level of hermetic sealing. The range of vacuum levels 
required ranges from $10^{2}$\,mbar in MEMS accelerometers to
$10^{-4}$\,mbar in microbolometers \cite{lindroos2009handbook}. Very low vacua are also needed in field emission
devices and the lowest recorded encapsulated pressure the authors have found in
the literature  ($10^{-8}$\,mbar) \cite{choi1999glass} was achieved using this technology. Maintaining
UHV is also important to photomultiplier tubes and we highlight the work of
Erjavec \cite{erjavec2001vacuum} who have performed a similar study to this one. Lower pressures in encapsulated micro-devices have probably been achieved,
but the means to measure them do not exist as most gauges with capability down 
to UHV have far greater internal volumes than the devices themselves. We are fortunate that the device we are aiming to
produce, by its very nature, is capable of measuring such low pressures. It is
commonly known in the atom trapping field that the loading rate, $\gamma$\,(Hz), of
an atom cloud is linearly related to the background pressure, with an
approximate scaling of $2\times10^{-8}\gamma$\,mbar\,s, and
Arpornthip et al  \cite{arpornthip2012vacuum} performed a systematic study
of this gauging technique. It was found to vary little with systematic variations, 
such as cooling beam power and detuning, and had a sensitivity range from $10^{-7}$
to below $10^{-9}$\,mbar, limited by cooled atom collisions within the trapped cloud. 
This sensitivity range is slightly above the range of our target pressure but will 
provide an adequate indication of the internal environment. An improved sensitivity 
down to $10^{-12}$\,mbar may be possible if the background rubidium vapour can be 
quickly reduced after loading \cite{willems1995creating} using the techniques discussed in the previous section.

\begin{figure}
\includegraphics[width=0.45\textwidth]{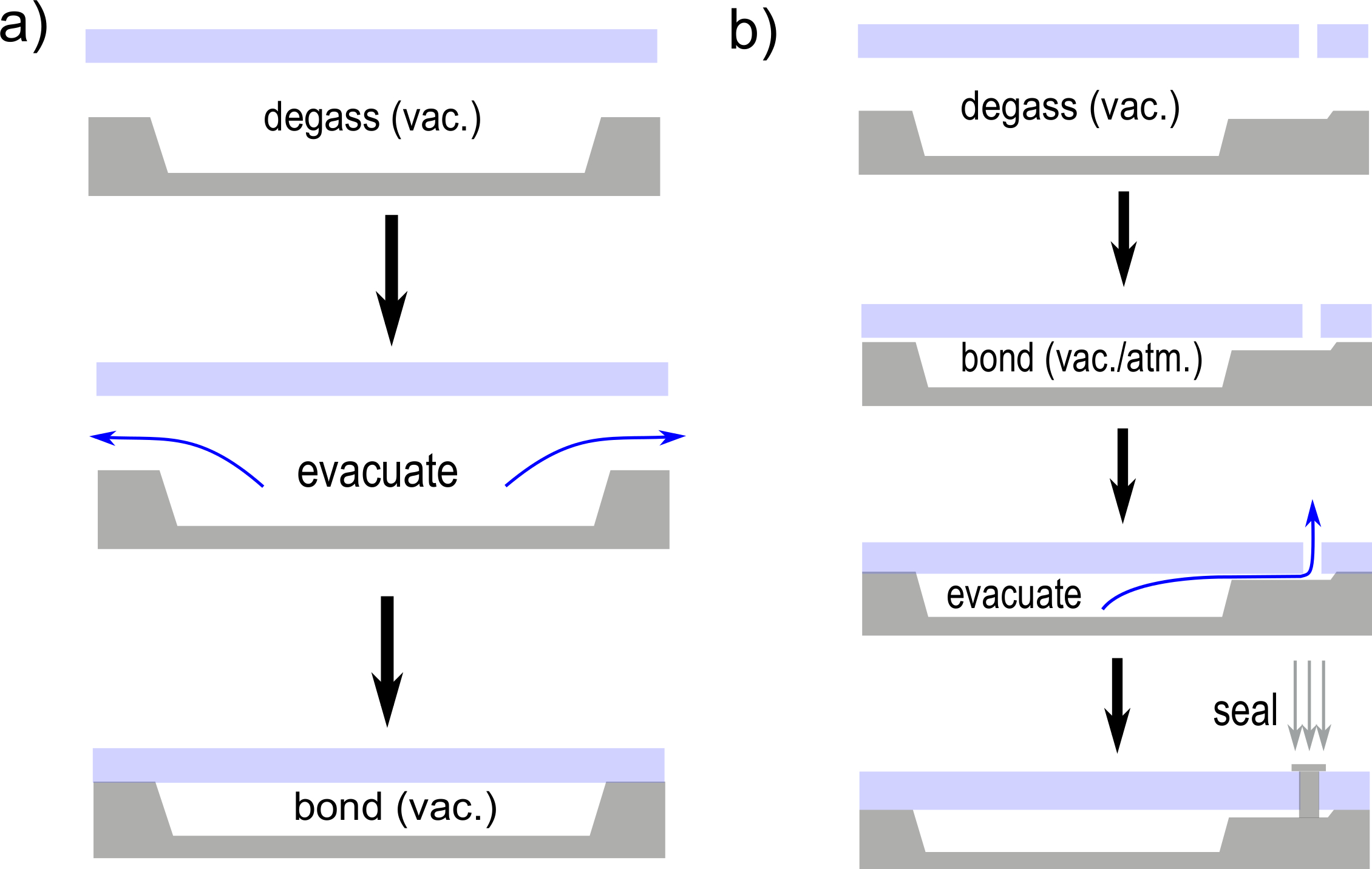}
\caption{\label{bond} Techniques to encapsulate a vacuum by evacuation before
(a) and after (b) bonding.}
\end{figure}

Two methods are commonly
used to encapsulate low pressures inside chips \cite{lindroos2009handbook}: wafer to wafer bonding under vacuum,
or sealing of an evacuation tube after bonding (as shown in Figure \ref{bond}. The
latter is simpler to construct as the various chip layers do not need
to be manipulated and bonded under UHV.
This method has been used in the
NIST atom chip system in which the evacuation tube is constantly pumped by a
miniature ion pump. However, efficient evacuation to UHV through a small aperture is
difficult and the sealing process is non-trivial on a wafer-level scale in terms
of complexity, uniformity and time. We shall see that
wafer-to-wafer bonding under vacuum, whilst complicated to implement, allows one
to thoroughly degas and evacuate chips and also allows the introduction of a
novel structural scheme to drastically reduce leak rates discussed in Section \ref{leaksec}.
\vspace{3mm}

The choice of materials from which the micro-MOT can be constructed is
dependent on the sealing method and the typical process environments. Suitable
materials for UHV have negligible vapour pressures, low outgassing and
permeability rates, and are mechanically strong and machinable. Therefore metals such as stainless steel, aluminium, titanium and copper are predominant. The MicroMOT would require optical access for the cooling beams and fluorescence detection and
so the chip must include an optically transparent section, such as a glass
or glass-ceramic wafer. Unfortunately there it a mismatch in the coefficients of thermal expansion 
(CTE) between glass and metal components which limit their baking, bonding, and operating temperatures. 
The exceptions are low expansion alloys such as Kovar and Invar,
the former being specifically developed for glass-metal seals. 
Any remaining CTE mismatch can typically be alleviated with specially shaped sealing edges \cite{roth1994vacuum}.
Nevertheless, we find that the majority of metals \footnote{The lowest outgassing rates for metals are tabulated by Redhead \cite{redhead1999ultimate} with high temperature annealed stainless 
steel and OFHC copper displaying outgassing rates between $10^{-15}$ and $10^{-16}$\,mbar\,l\,s$^{-1}$cm$^{-2}$.} 
do not have the extremely low outgassing rates \cite{perkins1973permeation,zhang2008outgassing} highlighted in Table \ref{uhvprop}, 
and their glass-metal seals \cite{roth1994vacuum} may not be adequately hermetic \footnote{To be more precise, the authors cannot  find absolute leak 
rate of these seals in the literature and commercial glass-metal, or glass-ceramic, seals only
state leak rates `below $10^{-10}$\,mbar\,l\,s$^{-1}$'. However, our discussions with commercial vacuum tube manufacturers
indicate that the leak rate for metal-ceramic seals could be as low as 10$^{-18}$\,mbar\,l\,s$^{-1}$ or lower.}, or are impractical to implement
into the microfabrication process (i.e. very high temperatures for extended periods). 
Alternative bonding methods are discussed in Section \ref{leaksec}.
\vspace{3mm}

The MicroMOTs are likely to incorporate atom chips which commonly use silicon as a substrate due to its high 
thermal conductivity and the vast array of available semiconductor processing
techniques \cite{reichel2010atom}. Coincidentally silicon, as we shall see in the following sections, is a very suitable UHV material: it has
extremely low permeation and outgassing rates at room temperature, it has several CTE-matched optical materials available, it is produced with a high purity (to the 9N
level), and can withstand high temperatures necessary for baking and bonding. 
The disadvantage of this material is its brittleness \cite{petersen1982silicon}, so only small structures can be fabricated using specialized techniques which is acceptable in our
application, but not for large scale vacuum systems.
Several well studied processes exist to clean silicon wafers and the lowest leak rates we have found for sealing technologies
have been found for silicon-glass bonding (Section \ref{leaksec}). It is also
interesting to note that polysilicon coatings are commercially available to
reduce the outgassing rates of stainless steel chambers
\footnote{http://www.silcotek.com/silcod-technologies/SilcoGuard-high-purity-coating/}.
Another advantage is that silicon is completely non-magnetic, which is important
for manipulating atoms, and is a poor electrical conductor which reduces the
deleterious effect of eddy currents during fast magnetic field switching \cite{dedman2001fast} - a common issue
with MOTs. Hence, in the following section we assume the chips are predominantly
constructed from silicon and glass, with additional metal films for reflectors
and getters.  

\subsection{Pumping}
\label{pumpsec}

One cannot maintain a high vacuum without any form of pumping because no seal is
perfect, all materials outgas to some extent, and no material is impermeable to
all gases. However, by reducing the above effects as much as possible one can
sustain vacuum with minimal pumping, especially in small volumes. As
mentioned in the introduction we would like to maintain UHV with no active or
cryogenic pumping as such systems increase the total size, power requirements, and
cost, not to mention the time and money needed to develop chip scale analogues of these devices.  Passive
pumping elements take the form of getters which are metals, or
alloys, that chemisorb typical gases found in high vacuum, namely O$_{2}$, CO,
N$_{2}$, and H$_{2}$. Getters generally come in two varieties \footnote{For a
more detailed up to date sub-division of getter types see Chuntonov et al \cite{a2013getter}.}: Evaporable getters are metals which are heated until their increased vapour pressure causes them to deposit on surrounding surfaces.  This traps residual gases under the deposited layers, but the new surface also acts as a pump to impinging gases through chemisorption.
Non-Evaporable Getters (NEGs) also chemisorb gases onto their surface, but in addition they absorb the reacted surface material into
the getter bulk during heating (known as activation). Both types keep pumping
gases at room temperature, albeit at a reduced rate, providing their surfaces are not saturated.
NEGs activate at temperatures ranging around 200-800$^\circ$C,
 depending on their composition, whereas evaporable getters need to be
heated to well above 700$^\circ$C. Therefore the choice of NEGs for our MicroMOT is obvious, 
and is further validated by its adoption in the MEMs vacuum encapsulation
industry.
\vspace{3mm}

Non-evaporable getters are made of Group IV/V metals and alloys, such as Ti, Zr,
V, Hf etc, and may also include metals such as Al and Fe. These are elements
with high oxygen solubility, high diffusivity, and high enthalpy of adsorption
for many gases found in vacuum \cite{benvenuti1999novel}. When exposed to air,
the surface of the NEG quickly passivates forming oxides, nitrides and carbides
in an 2-3\,nm layer \cite{sharma2008ti}. Heating the NEG in vacuum, known as
activation, causes these compounds to diffuse into the bulk leaving a fresh
metallic surface pump. Typical oxygen solubilities for NEG compounds are on the
order of 10\%, so a 1\,$\mu$m thin film can undergo $\sim$100 reactivation cycles
after air exposure, however the pumping efficiency begins to reduce after a few cycles \cite{chiggiato2006ti}. 
This corresponds to an approximate total
capacity of $10^{12}$ molecules per cubic centimetre. Recent \textit{in-situ} studies of NEG activation with individual gases at temperatures above activation indicate far higher capacities of the order of $10^{5}$ monolayers of carbon monoxide \cite{bender2010uhv} due to the increasingly uniform oxygen concentration in the film with temperature. Hydrogen
diffuses 
readily in the bulk, and so the capacity is approximately two orders of magnitude higher than surface pumped species at room temperature.
Embrittlement of the film at very high hydrogen concentrations (above 1\%) can result in
delamination and so should be avoided \cite{jousten2008handbook}. When NEGs are deposited as thin films they
also act as outgassing barriers \cite{benvenuti1998decreasing}, thus turning gas
sources into pumps and greatly reducing the ultimate pressure. Noble gases and
some hydrocarbons, such as methane, are not pumped by NEGs at room temperature.
Therefore the MicroMOT will need to be sealed at UHV to ensure the majority of
gases, especially the nobles, are evacuated prior to encapsulation.
\vspace{3mm}

The bonding techniques discussed in Section \ref{leaksec} require
temperatures up to 400$^\circ$C, which will lead to increased outgassing (see
Section \ref{outsec}) and a reduction of the NEG lifetime due to saturation. Moreover,
reactivation of the getter to rejuvenate its pumping rate during the sealed
devices' lifetime will lead to increased outgassing, permeation and possibly
leakage. Therefore it is prudent to use an NEG alloy with a very low activation
temperature, and high pumping rate and capacity. The beam lines of particle accelerators require XHV environments, and their
very large volume presents an issue for efficient and uniform pumping. Several
decades of research at CERN have been devoted to finding NEGs both which
activate during the chamber baking procedure ($\sim250^\circ$C) and can also
coat all internal surfaces \cite{prodromides2001lowering}. Their findings have
shown that sputtered TiZrV alloys of nearly equal ratios can be activated at
$180^\circ$C, and using these coatings they have demonstrated the 
lowest room temperature vacuum of $10^{-14}$\,mbar \cite{benvenuti2001ultimate}.
The pumping rate of NEG films depends on their surface area and so the CERN team
have also looked into the effects of substrate and deposition parameters to
increase pumping rates and capacities \cite{prodromides2002non}. Additional attractive properties of TiZrV NEG films
are their high adhesion, thermal and vibration stability, resilience to
standard wafer cleaning processes, and commercial availability \footnote{\textit{SAES getters} and \textit{Nanogetters}.}. Table
\ref{NEG} shows the typical pumping rates and capacities of TiZrV NEG thin films
whose values will be used in the following sections. Not all gases are pumped equally
and some lead to reduced pumping speed of the NEG at high surface coverage (for example, CO reduces the pumping rate and
capacity of H$_{2}$ and N$_{2}$) which are shown in Table \ref{NEG}. This effect must be accounted for when calculating the lifetime  of getter pumped devices. The unintentional incorporation of noble gases in sputtered films can result in outgassing which may endanger the vacuum \cite{amorosi2001study,window1993removing} and this will be discussed in Section \ref{outsec}. As a result, alternative methods, such as vacuum arc deposition  \cite{sharma2008ti} or e-beam evaporation, should be considered.
\vspace{3mm}

 Most gases only chemisorb on the NEG surface and show negligible pumping after
a monolayer is formed. Hydrogen is the exception as it diffuses throughout the
entire bulk of the getter and so only the thickness of the film defines the capacity.
There exists a thermal equilibrium between the absorption and desorption of
hydrogen from the NEG.  This is dependent on the hydrogen concentration \cite{liu2004kinetics} and thus can be used to
predict the residual pressure in our devices. This value, known as the
disassociation pressure, follows Sieverts’ law and has been measured for TiZrV
films \cite{chiggiato2006ti}.  It was found to be given by:
\begin{equation}
\log_{10}(P_{H_{2}})=2\log_{10}(x_{H})+14.324-\frac{8468}{T}
\end{equation}
Where $P_{H_{2}}$ is in millibar, $x_H$ is the fraction of hydrogen in the film and $T$ is the temperature in Kelvin. We
can see that for a very saturated film ($x_H=0.01$) the pressure is negligible at
room temperature ($10^{-19}$\,mbar) and only endangers the vacuum at temperatures above
150$^\circ$C, at which point helium permeation through the glass wafer becomes
equally problematic, as we shall see in Section \ref{permsec}.
\vspace{3mm}

Earlier we dismissed evaporable getters on the basis of high operating
temperatures and vapour phase gases. There is a new type of evaporable, or more specifically reactive, getter being investigated which uses alkali atoms as the gettering medium \cite{chuntonov2008new,chuntonov2008newb,chuntonov2008newc,chuntonov2009new}. Early studies of alkali metal dispensers showed
that they improve the pumping rate of the system \cite{Porta1968alkali}. Alkali and alkaline earth atoms will react and bind strongly to the common residual gases found at UHV and
therefore our devices may experience an improvement of the vacuum during operation. Experiments have shown that the pumping
rate for carbon monoxide by lithium getter films is similar to TiZrV NEGs but with a capacity
over $10^4$ times greater. This would be very advantageous to remove outgassed species
during bonding which could otherwise saturate TiZrV films. Most
of the work on these reactive getters has focused on lithium due to its low vapour pressure
and ability to form stable compounds with a number of gases. We suspect rubidium
will provide some gettering, but not to the extent of lithium due to the former's
high vapour pressure. For example, the disassociation pressure \cite{chuntonov2008new} of LiH at 300\,K is $10^{-21}$\,mbar whereas that of RbH is $10^{-7}$\,mbar, only marginally less than rubidium vapour pressure \cite{sangster1994h,steck2001rubidium}. It is expected, however, that pumping of oxygen and carbon monoxide will be more
effective. With its greater capacity and ability to pump additional gases such as methane, 
which NEGs cannot, a lithium getter could replace, or complement, the NEG
in the MicroMOT if the necessary vacuum cannot be pumped by NEGs alone\footnote{One could consider using lithium as the trapped species instead}.
  
\begin{table}[H]
\caption{\label{NEG}Typical pumping rates, sticking factors and capacities of
TiZrV NEGs at room temperature for a 1\,$\mu$m film. The values in parentheses indicate
the pumping rate after carbon monoxide saturation \cite{chiggiato2006ti}. The capacities can be increased
by nearly an order of magnitude by heating the substrate during deposition.}
\begin{tabular*}{0.45\textwidth}{@{\extracolsep{\fill} }lccc}
\toprule
Gas & Sticking & Pumping & Capacity  \\
 &  factor &  rate (l\,s$^{-1}$cm$^{-2}$) & (cm$^{-2}$)\\
\midrule
H$_{2}$ & $8\times10^{-3}$ & 0.35 (0.1) & $>10^{16}$\\
N$_{2}$ & $1.5\times10^{-2}$ & 0.17  (0.1) & $1.5\times10^{14}$\\
CO & 0.7 & 8  & $10^{15}$\\
\bottomrule
\end{tabular*}
\end{table}

In the following sections we assume:
\begin{itemize}
\item An internal volume of 0.5\,cm$^3$.
\item A surface area of 5\,cm$^2$.
\item  An NEG area of 1\,cm$^2$ with a thickness of 1\,$\mu$m.
\end{itemize}

\subsection{Permeation}
\label{permsec}

No material is absolutely impermeable to all gases, and so it is inevitable that
they will diffuse through the walls of any chamber and finally desorb into the
vacuum. We shall address the issue of permeation before the effect of outgassing
and leaks as this is seen by many as the limiting factor in achieving UHV in
small, sealed, well-degassed volumes. Permeation is complex and requires
several processes to become a vacuum risk: 1) The gas molecules in the external
environment impinge on the outer surface of the chamber wall and physisorbed, 2) they
disassociate if the surface enthalpy is greater than their bonds, 3) they
are absorbed under the surface layer and diffuse through the bulk
along the concentration gradient (Fick’s law), 4) the gas atoms must then overcome any
surface energy barriers, and 5) desorb from the surface
directly or recombine with other ions to desorb as a molecule. These processes 
strongly depend on the type of permeating gas and the chamber wall material. For
example, noble gases will permeate glasses, but not most metals due to the
latter's more crystalline structure and weak surface interaction.
As noble gases are not pumped by NEGs their permeation is of greatest importance and we shall devote the majority of this
section to them, however we will also discuss the effect of hydrogen as this is
the second fastest permeating gas (see Table \ref{gasperm}).
\vspace{3mm}

Surface effects, which will be discussed in Section \ref{outsec}, tend to reduce
the permeation rate so the simple process of bulk diffusion can be considered
the limiting factor of
permeation. The amount of gas flowing diffusively across a membrane of area $A$
and thickness $d$ between two regions of pressure $P_{ext}$ and $P_{int}$ is
\begin{equation}
\label{conperm}
\frac{dQ}{dt}=\frac{KA(P_{ext}-P_{int})}{d}
\end{equation}
where $K=DS$ is the permeation rate (cm$^{2}$\,s$^{-1}$), $D$ is the diffusion constant
(cm$^{2}$\,s$^{-1}$) and $S$ is the solubility (cm$^{3}$ (STP)/cm$^{3}$). Both $D$ and $S$ typically follow an Arrhenius-type temperature dependence and the former is quoted in the
literature as
\begin{equation}
\label{arrenheius}
D(T)=D_{0}\exp\left(\frac{-E_{D}}{k_{B}T}\right)
\end{equation}
where $E_{D}$ is the diffusion energy, $k_{B}$ is the Boltzmann constant and $T$ is the temperature. 
The variation of $S$ can be much more
complicated \cite{shackelford1972solubility}, however over a limited range of temperatures the Arrhenius 
form is adequate, and for most materials does not change appreciably 
compared to $D$ with temperature. Values of diffusion, solubility and permeability of the materials highlighted in this study 
can be found compared to the common vacuum materials, stainless steel and Pyrex, in Table \ref{diff}. Equation \ref{conperm}
assumes the gas is already fully dissolved in the membrane, which is not the case in thoroughly degassed materials.
 Therefore there is a period of time before the gas will `break through' to the evacuated volume, and in materials with very low
diffusion rates this can be extremely long \footnote{A common value quoted in the
literature is the `characteristic time' $t_c=\frac{d^{2}}{6D}$, this is time it
takes to produce a constant flow, not for the gas to initially
break through and so overestimates the time it take for the permeating gas to effect the vacuum.}. The pressure increase
of a cavity of volume $V$ at a time $t$ due to a gas permeating through a degassed membrane
is \cite{rogers1954diffusion}:

\begin{equation}
\label{permeation}
P_{c}=\frac{ADSP_{ext}}{Vd}\left[t-\frac{d^{2}}{6D}-\frac{2d^{2}}{\pi^{2}D
}\sum^{\infty}_{m=1}\frac{(-1)^{m}e^{\frac{-m^{2}\pi^{2}Dt}{d^{2}}}}{m^{2}}
\right]
\end{equation}

\begin{figure}[b]
\includegraphics[width=0.45\textwidth]{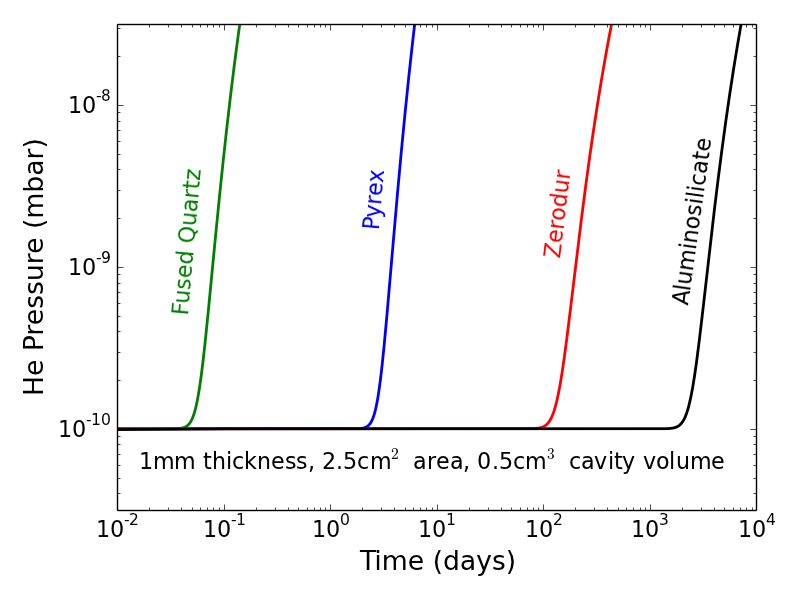}
\caption{\label{perm} The permeation rate of various 1\,mm optical wafers exposed to
atmospheric helium after initial evacuation to 10$^{-10}$\,mbar after complete degassing. 
We have calculated the values using Equation \ref{permeation} and used the data for \textit{Corning} 1720 to define the properties of AS glass \cite{altemose1961helium,covino1986laser}.}
\end{figure}

Since the micro-MOT chambers will require at least one optically transparent viewport 
the permeation of helium through glass will be a significant issue.
The amorphous network structure of glass forming oxides
provide channels for helium to diffuse. Not all
glasses are alike however, and the addition of `modifier' molecules can act to
plug the holes in the network, resulting in very low permeation rates \cite{altemose1961helium}. We aim to
seal glass to silicon and therefore must match CTEs to reduce stresses and therefore increase yield.
Pyrex is the most common
glass bonded to silicon due to their comparable CTEs and
its sodium content required for anodic bonding \cite{knowles2006anodic} (see Section \ref{leaksec}). This borosilicate
glass is so highly permeable to helium that we should expect to lose UHV 
 several days after bonding (see Figure \ref{perm}). Much
work was carried out in the 1960s and 1970s on the permeation rates of gases
through glass, and the results showed that Alumino-Silicate (AS) glass, those with
 approximately 20\% Al$_2$O$_3$ or more composition, had permeation rates five orders of magnitude
lower than Pyrex \cite{altemose1961helium,shelby1976helium}. Figure \ref{perm} shows a comparison of the permeation rates of
helium through AS glass, Pyrex, fused silica, and \textit{Schott} Zerodur glass-ceramic, calculated using Equation \ref{permeation}. 
This latter material is commonly used in UHV systems requiring low permeation and its very low CTE makes it well suited for bonding. We
can also see that \textit{Corning} 1720 series AS glass is more than capable of maintaining vacuum for our target time. Its CTE is well
matched with silicon to which is can be anodically bonded, 
albeit at much higher temperatures than Pyrex due to the low alkali content \cite{wallis1969field, baine1997single, spangler1988new}.

Other AS glasses are commercially available, but too numerous to list here, and we mention that load-borate and soda-lime glasses also have low permeability characterisics.
Silicon carbide \cite{sarro2000silicon} has a very low
permeability \cite{jung1992diffusion}, can be bonded to silicon \cite{tudryn2005characterization}, and have been used for atom chips
due to its transparency and high thermal conductivity
\cite{shevchenko2006trapping}. Sapphire and glass-ceramics such as Spinel \cite{trocellier2014review} are likely to have extremely
low permeation rates, but yet again are not well CTE matched. Hard crystalline optical
coatings may also reduce permeation. Graphene, amongst its many other attractive properties, has shown a
permeability rate equal to bulk Pyrex, yet requiring only a single monolayer \cite{bunch2008impermeable}. Unfortunately
uniform coatings over large areas are as yet unavailable, but graphene-oxide may
be a suitable alternative \cite{nair2012unimpeded}.
\vspace{3mm}

\begin{table*}
\caption{\label{diff} Bulk diffusivities and solubilities (at STP) of helium and hydrogen
in silicon, Pyrex (\textit{Corning} 7740), aluminosilicate (\textit{Corning} 1720), and
stainless steel. We have also tabulated the permeation rates at 20$^\circ$C
and 500$^\circ$C to compare values more easily.}
\begin{tabular*}{\textwidth}{@{\extracolsep{\fill} } lcccc}
\toprule
 & Silicon \cite{van1956permeation,alatalo1992first}&
Aluminosilicate \cite{altemose1961helium,souers1978permeation,steward1983review} 
& Pyrex \cite{altemose1961helium,souers1978permeation,steward1983review, kurita2002measurements, laska1969permeation, shackelford1972solubility} &
Stainless steel \cite{norton1957permeation,marin1998outgassing}\\
\midrule
\textbf{Helium\footnote{We are unable to find helium solubility in silicon data other than van Wieringen et al \cite{van1956permeation} which is somewhat unreliable due to the limited measurement range. In the tabulated values we have used that data with the theoretical energy of solution \cite{alatalo1992first} to calculate the prefactor. The diffusivity values are also theoretical but agree with experimental results \cite{Jung1994diff,van1956permeation,Griffioen1987nuc}.}} && & & \\
D$_{0}$  (cm$^2$s$^{-1}$)                 & $5.2\times10^{-3}$  & $3.7\times10^{-4}$   & $4.6\times10^{-4}$    & Impermeable \cite{norton1957permeation,chen2003diffusion} \\
E$_{D}$ (eV)                              & -0.82                & -0.52                 & -0.28                  & -\\
S$_{0}$  (cm$^{3}$\,(STP)\,cm$^{-3}$)       & $2.8\times10^{-4}$   & 0.0016   & 0.005   & - \\
E$_{S}$ (eV)                              & -0.77                & -                    & -                     & - \\
K  (cm$^2$s$^{-1}$) @ 20$^\circ$C& $3.2\times10^{-34}$ & $6.8\times10^{-16}$  & $ 3.5\times10^{-11}$ & -\\
\hspace{1.76cm}@ 500$^\circ$C            & $4.8\times10^{-17}$ & $2.4\times10^{-10}$ & $ 3.4\times10^{-8}$  & -\\
\midrule

\textbf{Hydrogen\footnote{We have assumed the same solubility for AS glass as for Pyrex due to the small variation found between glasses and the minimal temperature variation\cite{laska1969permeation,  kurita2002measurements}.}} 
\footnote{Like helium, the commonly quoted values measured by van Wieringen et al \cite{van1956permeation} are extrapolated from a narrow high temperature range and can be assumed to indicate the highest diffusion rate (see Section \ref{outsec}).} &&&&\\
D$_{0}$  (cm$^2$s$^{-1}$)         & $9.7\times10^{-3}$                 & $2.08\times10^{-7}$\,K$^{-1}$      & $1.4\times10^{-5}$                &  $1.2\times10^{-2}$\\
E$_{D}$ (eV)                      & -0.48                              & -0.67                             & -0.24                            & -0.56\\
S$_{0}$  (cm$^{3}$\,(STP)\,cm$^{-3}$)        & 90.4      & 0.038      &  0.038  & 0.3  \\
E$_{S}$ (eV)                      & -1.86                              & -0.12                                & -0.12                               & -0.11\\
K  (cm$^2$s$^{-1}$) @ 20$^\circ$C & $ 5.7\times10^{-40}$ &  $ 6.1\times10^{-23}$ & $ 3.4\times10^{-16}$ &  $1.0\times10^{-14}$\\
\hspace{1.76cm}@ 500$^\circ$C     & $ 1.2\times10^{-15}$  &  $ 4.3\times10^{-14}$ & $ 2.4\times10^{-12}$  &  $1.4\times10^{-7}$\\
\bottomrule
\end{tabular*}
\end{table*}

Norton \cite{norton1957permeation} measured the permeation of other gases
through fused silica, as shown in Table \ref{gasperm}, and we can see that, in general, 
larger molecules have lower permeation rates but can depend on surface interactions and solubilities (compare H$_2$ and Ne). 
Hydrogen diffuses through glass as a molecule \cite{kurita2002measurements, shackelford1972solubility}
and so, as shown in Table \ref{diff}, it will have a far lower permeation rate than helium. 
Coupled with an order of magnitude lower atmospheric partial pressure compared to helium, and the ability
to pump the gas with NEGs, hydrogen permeation can be neglected. Very little data is available on 
the hydrogen permeability through AS glass and so we have used a scaling law by Souers et al \cite{souers1978permeation}, 
the glass composition from Altemose \cite{altemose1961helium}, and the Pyrex solubility to calculate the diffusivity constants in Table \ref{diff}. 

\begin{table}[H]
\caption{\label{gasperm} Permeation of different gases \cite{mantina2009consistent,batsanov2001van} through fused silica \cite{norton1957permeation}.}
\begin{tabular*}{0.45\textwidth}{@{\extracolsep{\fill} }lcc}
\toprule
Gas & Relative permeation & Van der Waals  \\
 &  rate at 700$^\circ$C &  radius, nm  \\
\midrule
He & 1 & 0.133\\
H$_{2}$ & 0.1 & 0.15\\
Ne & 0.02 & 0.141\\
Ar & <$10^{-7}$ & 0.176\\
\bottomrule
\end{tabular*}
\end{table}

We now address the second material in our system: silicon. Measurements
of helium permeation through silicon extrapolated from high temperatures show
that, like metals, silicon is practically impermeable to all noble gases. Using
measured values for solubility cite{binns1993hydrogen,van1956permeation} and
 typical atmospheric helium content, we should not expect to find a single atom within a
 cubic centimetre of silicon. We note that recent studies \cite{sparks2013hermeticity, sparks2013output} looking at the hermeticity of glass frit encapsulation and other bonding methods have indicated that helium permeation through silicon
at room temperature may be more significant than expected. 
However, more work is required to confirm this against the large bulk of research into 
helium bubble formation in silicon which agrees with the original low permeation
result \cite{cerofolini2000hydrogen}. 
Hydrogen, on the other hand, is known to permeate silicon albeit predominantly in
atomic form at room temperature. This matter will be discussed in great detail in Section
\ref{outsec} as the permeation rate is related to outgassing. The results in
Table \ref{diff} show that the permeation rate of hydrogen through silicon at room
temperature is negligible.  
 
\subsection{Leaking}
\label{leaksec}

No seal is perfect as the bonding of materials will inevitably lead to a route for gases to
travel, via micro-channels and defects, or merely a local variation in
the permeation rate. Standard UHV systems
predominantly use Conflat type seals which employ knife-edges to bite into OHFC copper
gaskets and join metal components together, and `housekeeper' type seals for
glass-to-metal interfaces. Commercial vacuum products quote leak rates less than
$10^{-11}$mbar\,l\,s$^{-1}$ (STP), usually limited by the resolution of the leak
detector \footnote{Ultra low helium leak detectors using the accumulation
method and NEG pumps have measured leak rates \cite{bergquist1992innovations} as low as $10^{-15}$mbar\,l\,s$^{-1}$
.}. These types of seals are not suitable for
wafer-level fabrication of vacuum encapsulated micro-electronics and so several
new methods have been developed using chemically formed seals, or simply relying on the
attraction between perfectly flat surfaces \cite{christiansen2006wafer}. The small volume and long lifetime of many microelectronic chips may
preclude the use of internal pumping mechanisms, meaning that the seals must have extremely low leakage, more so than
those demanded by standard vacuum systems.

 In Section \ref{UHVsec} we discussed the methods to
encapsulate microfabricated vacuum devices and here we shall explore the details
of suitable bonding methods and their quality. We foresee the need for at
least two bonding processes: one to bond the `atom chip’ to the structure wafer
(silicon to silicon, or gold to silicon), and another to bond the glass capping
wafer to the structure layer (glass to silicon). As many bonding technologies
exist we will only consider those which are well established, have
demonstrated leak rates below $10^{-13}$\,mbar\,l\,s$^{-1}$ (air), and do not require
temperatures above 400$^\circ$C so as to reduce outgassing, stress, and protect chip
components. We have not considered low temperature indium
bonding which although initially seems promising can result in noble gas outgassing
unless special measures are undertaken. Also, it limits the activation of NEGs, may require
several additional films to improve surface wetting, and the
leak rate is not sufficiently low \cite{straessle2013low}.

\subsubsection*{Glass Frit}
Glass frit bonding is a well established technique which involves the deposition
of a low melting point glass compound between two materials. The glass is
heated first to outgas the organic binder compounds, and is then raised to the glass transition temperature which melts and seals the two surfaces upon
cooling. The vacuum hermeticity of this technique has been explored extensively
by Sparks et al \cite{sparks2004reliable}, but no absolute leaks rates have been quoted. We believe glass frit bonding to have a leak rate below $10^{-15}$\,mbar\,l\,s$^{-1}$ by
considering the lifetime, internal volume, and pressure inferred by the
integrated resonator's Q-factor \cite{sparks2006all}. However, the
pressure measured ($\sim10^{-3}$\,mbar) is at the limit of the gauging technique and one does not know the residual pressure immediately after bonding. Possible issues may include insufficient degassing of the organic binder materials, limitations on pre-baking  temperatures, incompatible CTEs (although they can be engineered to match the application) and the need for additional materials. Of the four bonding methods presented here, glass frit has the least sensitivity to surface quality and can be used to seal electrical feedthroughs.
\subsubsection*{Eutectic Bonding}
A eutectic alloy is one where the melting point of the constituent materials is
lowered on contact. For example gold and silicon individually have
melting points above 1000$^\circ$C, but when they are pressed together they will melt at
363$^\circ$C at their interface \cite{wolffenbuttel1994low, lani2006gold}. Subsequent cooling will form an alloy with high hermeticity and a strong bond. Other eutectic alloys exist, such as gold and tin \cite{lee1993bonding}, but we
highlight the gold-silicon system as many atoms chips employ gold as a reflector and conductor. This bond has demonstrated the lowest leak rate that we have found \cite{lindroos2009handbook}, below $10^{-15}$\,mbar\,l\,s$^{-1}$. Pssible disadvantages include the need for multilayer films to prevent interlayer diffusion \cite{lani2006gold}, the requirement of inert gas storage before bonding to prevent the native oxide growth on silicon, and the temperature restrictions post bonding, as further heating remelts the alloy and degrades the bond.

\subsubsection*{Anodic Bonding}
Anodic bonding occurs between an oxide forming metal (or semiconductor) and an alkali containing glass, by heating the two materials together (300-500$^\circ$C) with
the simultaneous application of a high voltage (100-1000\,V) across the interface\cite{knowles2006anodic, wallis1969field}. The mobility of the
alakli ions (typically sodium or lithium) in the glass is increased with temperature and they are pulled away from the interface by the electric potential. The residual non-bridging oxygen atoms at the interface then bond with the silicon. The high electric
potential gradient has the additional effect of pulling the two surfaces into
intimate contact which overcomes surface inhomogeneities. The two materials (usually silicon and borosilicate glass) must have very flat surfaces, below 10\,nm, and be CTE matched to avoid stress fractures during cooling. Hermeticity measurements show that the leak rate is below $10^{-14}$\,mbar\,l\,s$^{-1}$, with few residual gases other than oxygen, which is produced during bonding at the inner seam. This residual gas source can
be significant, especially in small evacuated volumes, and so getter films are 
mandatory (see Section \ref{outsec}). The leak rate measurements \cite{mack1997analysis, mack1997gas} included the effect of the bonding area around the cavity and found no variation, from which we infer that the seal is absolutely hermetic and possibly limited only by permeation.

\subsubsection*{Direct Bonding}
Direct bonding is the result of the attractive Van der Walls forces
between atomically flat surfaces. Semiconductor and glass wafers are routinely
produced with the required flatness and this technique requires no additional materials, 
does not release gases (unlike anodic and frit bonding), and is hermetic \cite{mack1997gas} (Leak rates below $10^{-14}$\,mbar\,l\,s$^{-1}$). Unfortunately, for the silicon-silicon direct bond, very high post-annealing temperatures (up to 1000$^\circ$C) are required to ensure a high bonding yield and to reduce voids. This latter effect is due to gases desorbing between the surfaces which becomes trapped. Bonding in vacuum produces fewer voids, due to a thorough degassing before sealing, to the point that strong bonds form at room temperature without the need for post annealing \cite{gosele1995self}. The extremely flat surfaces are difficult to retain during processing and even sub-micron particles will
result in debonding. CTE-matched glass-silicon direct bonding occurs with lower
annealing temperatures \cite{chuang2010study}, and the glass can absorb gases released between the interface.
\vspace{3mm}

To summarize the bonding techniques, there are several methods to obtain
reliable leak rates lower than  $10^{-14}$\,mbar\,l\,s$^{-1}$ and even below
$10^{-15}$\,mbar\,l\,s$^{-1}$. Anodic and direct bonds may even be absolutely hermetic, limited by permeation,
but their absolute leak rates were beyond the sensitivity of their measurements.
Eutectic bonding has shown the lowest measured leak rates and is a reliable and
low outgassing method. Glass frit may also have equally low leak rates, but
the residual gas pressure released during bonding is unknown.
\vspace{3mm}

\begin{equation}
\label{vaceq}
Q_{L}=\frac{V\Delta P}{\Delta t}
\end{equation}

Using Equation \ref{vaceq} we can calculate the highest permissible leak rate
for a gas at atmospheric pressure leaking into our specified volume of $V=0.5$\,cm$^{3}$, such that
the pressure does not rise by 50\% ($\Delta P=0.5\times10^{-10}$\,mbar) over $\Delta t=1000$\,days, to be
$3\times10^{-22}$\,mbar\,l\,s$^{-1}$. This rate seems unachievable, but we must
consider that some gases are pumped away by the NEGs and so may permit a higher
leak rate, and many gases have low atmospheric partial pressures. For example
the partial pressure of helium in the atmosphere is $5\times10^{-3}$\,mbar and so one can
permit a leak rate \footnote{The relative conductance, or leak rate, of non-interacting gases, 
is proportional to the square root of their masses. Hence helium will leak 2.6 times faster 
than nitrogen. For the following calculation we has assumed the average molecular mass of is 29.} of  $1.5\times10^{-17}$\,mbar\,l\,s$^{-1}$. \\
Reactive gas leakage is limited by the pump rate and capacity of the NEG films. In light of this we may model the lifetime of the device due to NEG saturation using the following
formula:

\begin{equation}
\label{getterlifefull}
\frac{dP_{c}}{dt}=\frac{1}{V}\left(Q_{L}-L_{P}P_{c}(1-\theta)^k\right)
\end{equation}

Where $P_{c}$ is the internal pressure (mbar), $Q_{L}$ is the leak rate (mbar\,l\,s$^{-1}$), $L_P$ is the pumping rate of the getter (l\,s$^{-1}$), $\theta$ is the fractional surface coverage, and $k$ is the order of desorption. 
The effect of surface coverage on pumping depends on the gas/surface chemistry, temperature, and surface geometry (flat, granular,
etc.). This topic is too extensive to detail here but can be found in most surface science graduate texts \cite{attard1998surfaces}. 
We find that the Langmuir adsorption isotherm, $(1-\theta)^k$, models the majority of data on NEG pumping rates \cite{chiggiato2006ti} 
adequately for our purposes due to the low surface coverage. 
The pressure inside the cavity will drop to base value:

\begin{equation}
\label{base}
P_{base}=\frac{Q_{L}}{L_{p}}
\end{equation} 

As noted earlier, the effect of saturation by some gases (for
example carbon monoxide) reduces the pumping speed of other gases \cite{chiggiato2006ti}
which we do not take into account dynamically in the model, but assume the
lowest pumping speed as the `worst case scenario'. 

\begin{figure}
\includegraphics[width=0.45\textwidth]{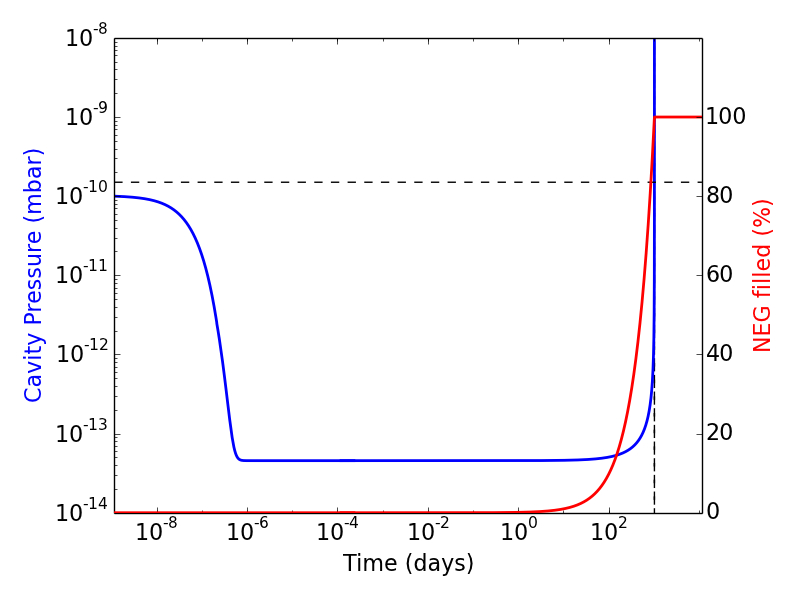}
\caption{\label{getter} The internal pressure (blue) and NEG 
coverage (red) with an atmospheric leak of nitrogen at a rate $6.5{\times}10^{-15}$\,mbar\,l\,$s^{-1}$ (air)
calculated using Equation \ref{getterlifefull}. The horizontal dashed line indicates the target pressure and
the vertical line the result of Equation \ref{getterlifeshort}. The cavity is initially at $10^{-10}$\,mbar. }
\end{figure}

There is a sharp drop in pumping speed at saturation allowing us to 
simplify the lifetime calculation. By assuming that the pumping rate is constant until the NEG has reached its capacity of $C_{G}$\,(moles), at which point it ubruptly drops to zero, and by setting $\theta=1$ in Equation \ref{getterlifefull} and using the ideal gas formula,
we find an approximate value for the maximum permissible leak rate: 
  
\begin{equation}
\label{getterlifeshort}
Q_{L}(max)=\frac{C_{G}RT}{\tau_{L}}
\end{equation}

where $R$ is the ideal gas constant, $T$\,(K) is the temperature and $\tau_{L}$\,(sec) is the lifetime. 
If we assume a pumping speed for nitrogen, the most abundant atmospheric gas, of 0.1\,l\,s$^{-1}$, 
a capacity of $10^{14}$ molecules,
with a lifetime of 1000 days: the maximum
permissible leak rate is $5{\times}10^{-15}$\,mbar\,l\,s$^{-1}$ with $P_{base}=5{\times}10^{-14}$\,mbar.
This can be achieved with reliable bonding from all the methods detailed earlier.   
Note that Equation \ref{getterlifeshort} is independent of both the pumping rate and background pressure because it assumes them to be at equilibrium, therefore one must use this equation alongside Equation \ref{base} to
ensure the correct base pressure. Figure \ref{getter} compares the numerical solution of Equation \ref{getterlifefull} with the approximate value from Equation \ref{getterlifeshort} and we find perfect agreement.
We note that the capacity is that for a single monolayer and so the lifetime can simply be extended through reactivation cycles. Hydrogen diffuses into the bulk and so reactivation does not increase the getter lifetime, but the NEG capacity for hydrogen can be two to three orders of magnitude greater (see Table \ref{NEG}), 
so is not as much of a concern. 
\vspace{3mm}

The atmosphere contains several noble gases \cite{lindroos2009handbook} including argon (9.3\,mbar), neon
($1.8{\times}10^{-2}$\,mbar), and helium ($5{\times}10^{-3}$\,mbar), where the values in parentheses are the
atmospheric partial pressures. Their leak rates are proportional to $(T/M)^{0.5}$,
where $T$ is the temperature in Kelvin and $M$ is their mass, hence helium leaks at
the fastest rate and is often used in hermeticity tests \cite{millar2009mems} known as `Helium Bombing'. If we assume a leak rate which is proportional to the
pressure differential across the bond, the effect of argon, due to its relatively high
atmospheric partial pressure will be most significant. The maximum permissible leak 
rate for argon is $1{\times}10^{-19}$\,mbar\,l\,s$^{-1}$. The significance of this gas has been mentioned in the literature \cite{reinert2005assessment}, but is generally ignored as helium permeation through glass is considered to be a more pressing issue.
If we compare the lowest measured leak rate for the bonding methods
of $10^{-15}$\,mbar\,l\,s$^{-1}$ (air) we find all the noble gases endanger UHV, however helium and neon leakage need only be reduced by factor of 15 and 25, respectively, which
may be possible with thicker bonding seams or external barrier coatings, whereas argon must be reduced by nearly $10^4$.

\begin{figure}
\includegraphics[width=0.45\textwidth]{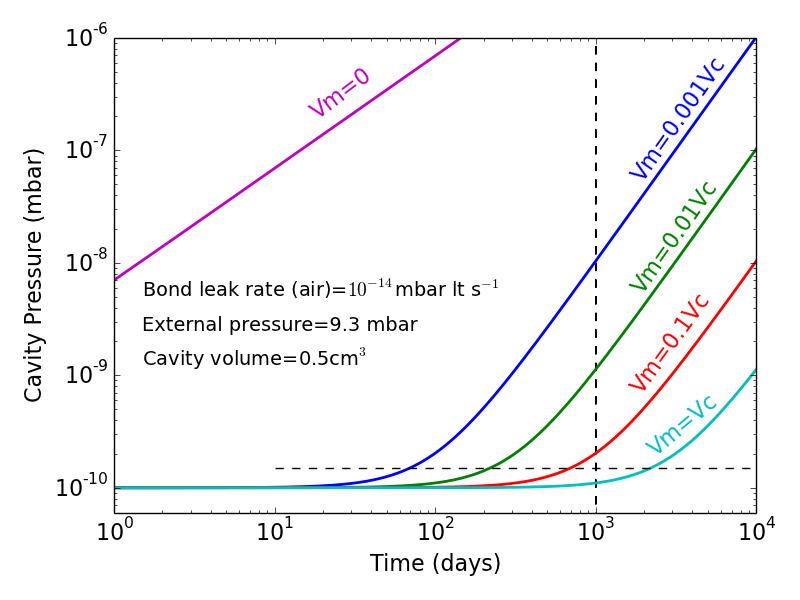}
\caption{\label{moat}Numerical solution of Equations \ref{moateqn} for various
ratios of moat/cavity volume. We assume argon is leaking from the
atmosphere, and the quoted leak rate is for each bond seam (air-to-moat and
moat-to-cavity) adjusted to atmospheric pressure to compare with literature
values. The horizontal dotted line represents a 50\% rise in pressure from an
initial value of $10^{-10}$\,Pa, and the vertical line indicates 1000 days. In this
simulation $V_{m}=0.25\,V_c$ would meet the target property with an overall leak rate
of  $6{\times}10^{-21}$\,mbar\,m$^{3}$s$^{-1}$ (air).}
\end{figure}

There is, however, a very simple scheme to reduce the
leak rate by several orders of magnitude: Simply by placing the vacuum chamber inside another. This can be achieved practically by introducing a buffer cavity, or moat,
within the seam such that the slow leakage into the moat results in an even
slower leakage into the main vacuum cavity \cite{gan2009getter}. This can be modeled by the following formulae and solved numerically:
\begin{equation}
\label{moateqn}
\frac{dP_{b}}{dt}=\frac{1}{V_{b}}\left(C_{ab}(P_{a}-P_{b})-C_{bc}(P_{b}-P_{c}
)\right)
\end{equation}
\begin{equation}
\frac{dP_{c}}{dt}=\frac{C_{bc}(P_{b}-P_{c})}{V_{c}}
\end{equation}
where $P_{i}$ is the pressure, $V_{i}$ is the volume, and $C_{ij}$ is the
conductance between $i$, and $j$, in which the subscripts $i,j=a,b,c$ refer to
the air, buffer, and cavity, respectively. We have independently modelled this effect and found stark, but advantageous, differences from the original study by Gan et al \cite{gan2009getter}. We suspect that an error was made in
tabulating their results, which also clarifies their unexplained lifetime
increase for 100\,mbar cavities. If we assume a main cavity volume of 0.5\,cm$^3$ and a
moat volume of 0.05\,cm$^3$ we can reduce the leakage rate by factor of $2{\times}10^{5}$ as shown in Figure \ref{moat}. This allows us
to use bonds with leak rates in the range of $10^{-14}$\,mbar\,l\,s$^{-1}$ (air) which is
technically feasible with all the bonding techniques considered earlier, and also
reduces the NEG limitations on reactive gases. The moat does not have to be
bonded at UHV as the model shows very little variation below an initial moat
pressure of $10^{-6}$\,mbar.

\subsection{Outgassing}
\label{outsec}


Outgassing is the release of gas from a material's surface and bulk. We
distinguish this from permeation in that the latter assumes a completely
degassed material, whereas outgassing is the result of gases that are difficult
to remove from the bulk.  These have either entered from diffusion during storage,
processing, or from the production of the material itself.
Cleanliness is of utmost importance in achieving UHV and we assume all
components have been through standard wafer cleaning processes, such as an RCA and ozone plasma. This latter technique has also been shown to improve bond strengths and reduce temperatures direct and anodic bonding \cite{choi2002analysis, zucker1993application}.  We shall not go into further detail regarding cleaning here
and direct the interested reader to the references \cite{kern1990evolution, kern1993handbook, vig1985uv, franssila2010introduction}.
We do highlight that detergents were found to remove vanadium from the NEG films and
so should be avoided \cite{taborelli2006cleaning}. 
\vspace{3mm}

Outgassing is the largest source of gas in well-sealed UHV systems and so usually
defines the lowest base pressure for a specific pumping rate $L_{p}$ according
to Equation \ref{base}. Hydrogen is the dominant gas at UHV, for which the NEGs have
a pumping rate on the order of 0.1\,l\,s$^{-1}$cm$^{-2}$, and so to achieve $10^{-10}$ mbar
one must ensure the outgassing rate is below $10^{-11}$mbar\,l\,s$^{-1}$. There is no
standard model for predicting the outgassing properties of all materials as many
different mechanisms are involved \cite{elsey1975outgassing1,elsey1975outgassing2} but can be essentially split into two
sources: surface and bulk. We assume that the surfaces are
clean in that common contaminates such as organics have been
thoroughly removed leaving only atmospheric and some processing species, namely
water, hydrogen, carbon monoxide, and noble gases. We can calculate the surface desorption rate using:
\begin{equation}
\label{desorb}
\frac{dN}{dt}=\frac{N\theta^{k}}{\tau_{s}}\exp\left(\frac{-E_{S}}{k_BT}\right)
\end{equation}
where $N$ is the surface density (molecules\,cm$^{-2}$), $\theta$ is the fractional surface coverage, $k$ is the desorption order,  
$\tau_s$ is the sojourn time (typically $10^{-13}$\,s), and $E_{S}$ is the desorption energy \footnote{typically $k=1$ for high surface coverage,
but increases to 2 when desorption is recombination limited at low coverage.}.
Typically, outgassing from the surface occurs at the fastest rate as physisorbed, or weakly chemisorbed 
gases, have low desorption energies.  Strongly bound molecules ($E_S$>
1.1\,eV) can in general be ignored as they do not appreciably desorb from surfaces
at room temperature, whereas weakly bound species ($E_S$< 0.7\,eV) can be pumped
away quickly. Molecules in the middle of this range are difficult to pump out
in a practical time and so high temperature baking is required \cite{lewin1965fundamentals}.
We shall refer to this as the `outgassing energies range (OER)'.
\vspace{3mm}

Outgassing of reactive species from the bulk is more complex than simple
diffusion. Gases, such as hydrogen, diffuse ionically and can form bonds with
the bulk material or impurities in a process known as trapping. Ions may also recombine
within the solid and become trapped in lattice defects, and any ion reaching the
surface needs to recombine in order to desorb. At low surface coverage this
latter, second order, step can be the limiting rate. Many of the transport
processes are activated and so only occur at elevated temperatures. This can
result in unreliable predictions when extrapolating high temperature data down to
room temperature. The effects are further complicated by surface oxides or
nitrides which, in general, act to reduce outgassing rates by providing a
barrier layer \cite{perkins1971diffusion, nemanivc2012hydrogen, nickel1995hydrogen}. Noble gases, on the other hand, only travel diffusively through
the bulk and easily desorb from surfaces at all temperatures due to their weak
interaction.  All of the effects outlined above act to only reduce the outgassing
rate compared to a simple diffusion model and therefore one can assume bulk diffusion as the most significant factor.  
If we assume purely diffusive outgassing from the material
bulk, and that it is degassed from both sides, then we can use the rate calculated by Lewin \cite{lewin1965fundamentals} for a `slab' geometry:

\begin{equation}
\label{degaseq}
\frac{Q_{OG}}{A}=\frac{8x_{0}D}{d}\sum_{m=0}^{\infty}\exp\left[-Dt\left(\frac{
\pi(2m+1)}{d}\right)^{2}\right]
\end{equation}

Where $D$ is diffusion rate, $A$ is the
surface area, $d$ is the thickness and $x_{0}$ is the initial concentration of the gas in the bulk. For a
non-disassociative gas $x_{0}=SP$, where $S$ is the solubility and $P$ is the
partial pressure of the gas. For a disassociative gas the concentration is
proportional to $P^{0.5}$ instead, and the solubility units are adjusted
accordingly. By using values for diffusion found experimentally, effects such as trapping 
are automatically included into the model.  
\vspace{3mm}
As highlighted by Chuntonov et al \cite{a2013getter} the increase of outgassing during the high temperature bonding process can cause the NEG film to become saturated and limit
the lowest obtainable vacuum and lifetime of the device. To calculate actual
lifetime including the effect of bonding we can consider the
reduction of getter capacity by the number of molecules released during bonding,
and using Equation \ref{getterlifeshort} to find:

\begin{equation}
\label{negbond}
\tau_{L}=\frac{T}{Q}\left(C_{G}R-\frac{\tau_{B}}{T_{B}}\sum_{gases}Q_{B}\right)
\end{equation}

where $T$ is room (or operating) temperature, $T_{B}$ is the bonding
temperature, $\tau_{B}$ is the bonding time (seconds), $Q$ is the outgassing
rate at $T$, and $Q_{B}$ is the outgassing rate at $T_{B}$. We have assumed the
temperatures are changed instantaneously and the bonding period is short enough
not to affect the operating outgassing rate. The same formula can be used to 
predict the increased outgassing due to reactivation of the NEG during the MicroMOT lifetime. The effect of the bonding can be
neglected if:

\begin{equation}
 \eta=\frac{\tau_{B}Q_{B}}{RC_{G}T_{B}}<<1
\end{equation}

Assuming an NEG with a hydrogen capacity of $10^{-7}$ moles ($x_{H}=0.01$), a
bonding time of $\tau_{B}=3600$s at a temperature of $T_{B}=400^\circ$C, and
we wish to keep $\eta=0.1$, the bonding outgassing rate must be
$Q_B<10^{-7}$\,mbar\,l\,s$^{-1}$cm$^{-2}$. If this is the result of bulk diffusion which 
scales as
\begin{equation}
\label{bakediff}
Q=\frac{D (T)}{D (T_{B})}Q_{B}=Q_{B}\exp\left(\frac{-E_{D}(T-T_B)}{k_BT_{B}T}\right)
\end{equation}
and we use a diffusion energy in the middle of the OER of $E_{D}=0.9$\,eV, then one must reduce room
temperature hydrogen outgassing rate to
$Q_B<10^{-16}$\,mbar\,l\,s$^{-1}$cm$^{-2}$. Equation \ref{bakediff} also applies for surface desorption (for which $E_D=E_S$). 
Gases such as carbon monoxide, for which the NEG has only a single monolayer capacity, require outgassing rates
over a hundred times lower compared to hydrogen without continuous reactivation. These are extremely low outgassing rates and are
the main hurdle in obtaining very low vacua in microelectronic devices. In the
following subsection we explore the outgassing rates of the main gases found at
UHV - H$_{2}$, CO, as well as noble gases - from the materials
considered for the MicroMOT and we have tabulated measured and theoretical
values for outgassing rates in Table \ref{lowout}.

\subsubsection*{Hydrogen}

\begin{figure*}
\includegraphics[width=\textwidth]{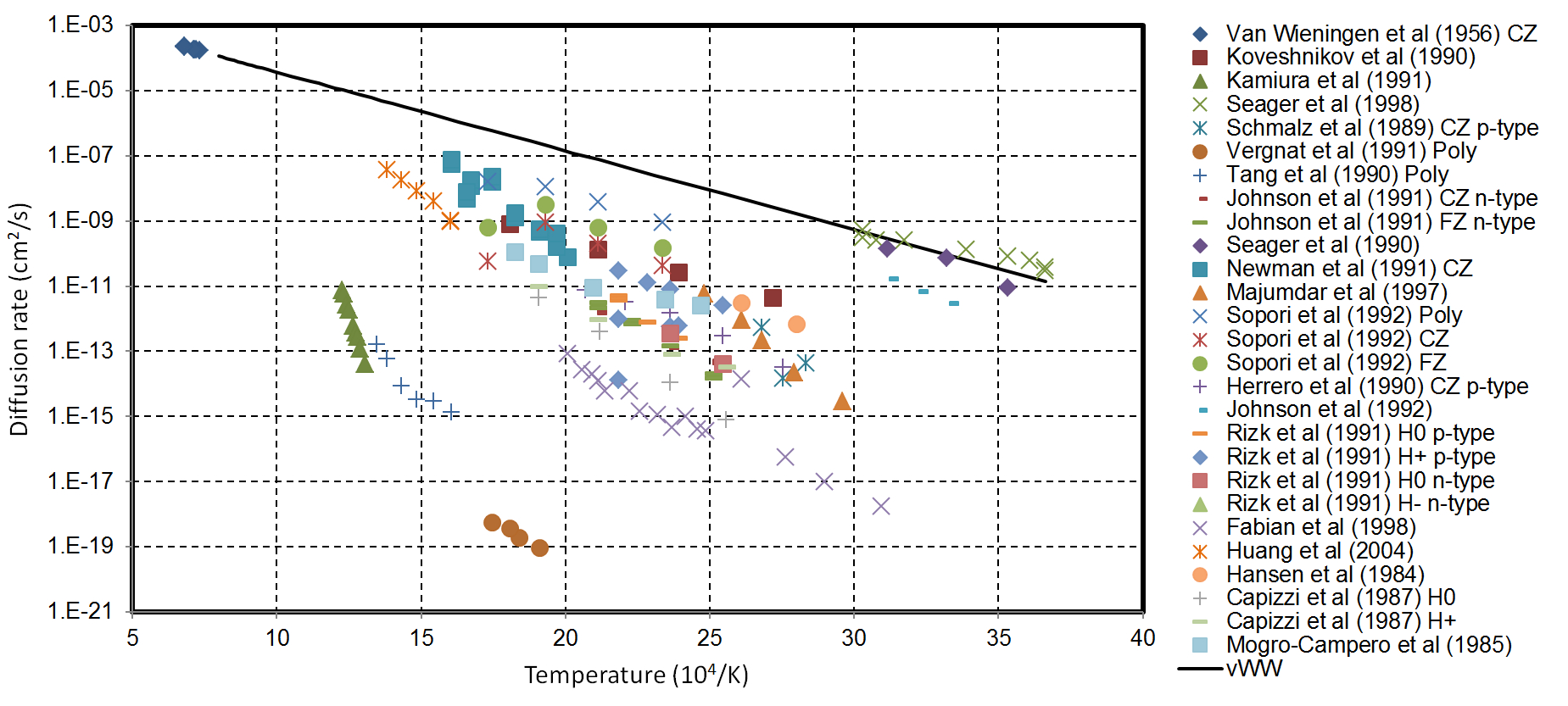}
\caption{\label{h2diff}The large scatter in data for diffusion of hydrogen
through silicon \cite{van1956permeation, koveshnikov1990peculiarities, 
kamiura1991hydrogen, seager1988real, schmalz1989hydrogen, vergnat1991determination, 
tang1990annealing, johnson1991stability, seager1990situ, newman1991hydrogen, 
majumdar1997reactivation, sopori1996hydrogen, herrero1990trap, johnson1992diffusion,
rizk1991hydrogen, fabian19981, huang2004hydrogen, hansen1984bulk, capizzi1987hydrogen, mogro1985drastic}. We have only plotted data for
hydrogen diffusion (no isotopes) and have indicated the type of silicon where
known. The solid black line is the commonly quoted vWW diffusion rate.}
\end{figure*} 

Hydrogen can be a major issue in semiconductor devices and so extensive work
has been carried out to understand its interaction with silicon \cite{pearton1987hydrogen, chevallier1988hydrogen,haller1991hydrogen,myers1992hydrogen,van2006hydrogen}. The most
common and earliest citation in the literature is to the work done by van Wieringen and Warmoltz \cite{van1956permeation} (which we
shall refer to vWW). Their diffusivity and solubility data was taken at very high
temperatures (967-1207$^\circ$C) and these values are
shown in Table \ref{diff}. Extrapolation down to room temperature is fraught with possible errors as hydrogen readily dissociates upon diffusing into the bulk, interacting with the silicon lattice and impurities in various forms.
Figure \ref{h2diff} shows the results of subsequent diffusion studies and one can see the large variation in measurements in the literature.
Advances in understanding have shown that hydrogen migration through bulk silicon predominantly occurs in 
atomic form at room temperature with an activation barrier of $\sim0.5$\,eV, but can recombine into a dimer 
which is then trapped by an barrier of $0.8-1.2$\,eV. Exact values are difficult to predict and depend on various doping and
impurity levels, growth methods, defects and lattice orientations, and even quantum effects \cite{miyake1998quantum}. 
The vWW diffusivity results are consistently higher than every subsequent measurement and can be taken as the upper limit in our analysis \cite{sopori2001silicon}. 

Whichever diffusivity one uses, the vWW solubility results extrapolate to $10^{-10}$ molecules per cubic 
centimetre at room temperature under 1\,bar of H$_2$ and so there should be no hydrogen whatsoever within the bulk. 
Other measurements have found agreement with a very low value , with 
the highest at only a few hundred hydrogen atoms per cubic centimetre at room temperature \cite{binns1993hydrogen, ichimiya1968solubility, mcquaid1991concentration}.
Using any one of these solubilities does not alter the permeation values in Table \ref{diff} by a more than a factor of two or three.

\begin{figure}
\includegraphics[width=0.45\textwidth]{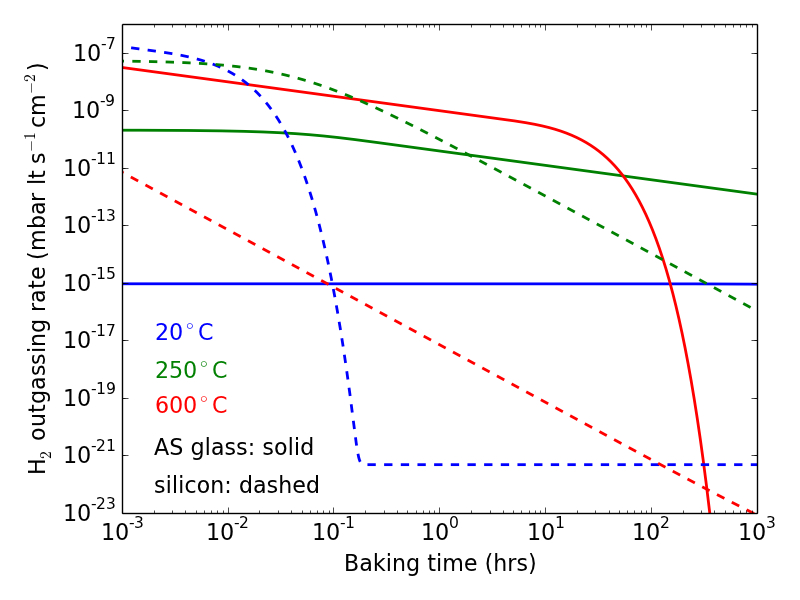}
\caption{\label{h2bake} The theoretical outgassing rate of hydrogen for 1\,mm thick silicon (dashed line) and AS glass (solid line) for three baking temperatures. We have assumed AS glass is diffusion limited and calculated the rate using Equation \ref{degaseq} and values from Table \ref{diff} after exposure to 1\,bar hydrogen (i.e. immersion into water). We have calculated the rate for silicon using a summation of diffusion (Equation \ref{degaseq}) from the bulk as well as considering the higher concentration at the surface (see text), and also recombination-limited surface desorption using Equation \ref{desorb} ($k=2$) with values from Gupta et al \cite{gupta1988hydrogen}. The room temperature silicon outgassing shows an initially high rate due to diffusion of the high concentration near the surface and is eventually limited by surface desorption of the dihydride surface species (as are the higher temperature bakes).}
\end{figure}

Standard semiconductor processing, such as mechanical polishing, 
HF etching, plasma treatments etc, can result in far higher levels of hydrogen close to the surface \cite{pearton1987hydrogen}.
Several studies have found values as high as $10^{18}$ molecules per cubic centimetre and drops significantly after a depth 
of one micron\footnote{Even higher surface concentrations can be produced with high energy ion bombardment, but we regard this as a non-standard process.}.
This concentration will lead to 
outgassing rates of $10^{-7}$\,mbar\,l\,s$^{-1}$cm$^{-2}$, using Equation \ref{degaseq} and the vWW diffusivity scaling law, but can be completely degassed to below $10^{-30}$\,mbar\,l\,s$^{-1}$cm$^{-2}$ within an hour under vacuum as shown in Figure \ref{h2bake}, limited by surface recombination. Moreover, surface oxides and nitrides act as efficient permeation barriers \cite{perkins1971diffusion,nickel1995hydrogen,nemanivc2012hydrogen}. Hydrogen bound on the silicon surface has very high desorption energy \cite{durr2006dissociative, gupta1988hydrogen} (above 1.8\,eV) such that the desorption 
rate from one monolayer coverage would be below $10^{-22}$\,mbar\,l\,s$^{-1}$cm$^{-2}$ using Equation \ref{desorb}. Thermal desorption studies \cite{gupta1988hydrogen} show that most hydrogen complexes can be desorbed from silicon by annealing at 600$^\circ$C.
\vspace{3mm}

Hickmott \cite{hickmott1960interaction} studied the interaction between hydrogen and glass and found that the hot filament of the ionization gauge had a detrimental effect on determining the residual gas content at UHV. He noted that hydrogen was desorbed at the two distinct
activation energies of 0.29\,eV and 1.08\,eV. The former is so low that it will
desorb completely at room temperature under vacuum, whereas the latter requires
baking above 400$^\circ$C. Spectroscopic studies by Hickmott showed that after a
high temperature bake the main residual gases were water and carbon monoxide. Todd \cite{todd1955outgassing} measured the residual water composition in a variety of glasses and found negligible outgassing ($\sim10^{-23}$\,mbar\,l\,s$^{-1}$cm$^{-2}$) in AS glass after high temperature baking. This low
outgassing rate is due to the strong Si-H and Si-OH bonds. Using the values from Table \ref{diff} and Equation
\ref{degaseq} to calculate the lowest hydrogen outgassing rate from AS glass, as
shown in Table \ref{lowout}, we find remarkable agreement between the theoretical
value of just over $10^{-17}$\,mbar\,l\,s$^{-1}$cm$^{-2}$ and experimental result \cite{hobson1964measurements,redhead1999ultimate} of just below\footnote{Hobson \cite{hobson1964measurements} measured a `leak rate' (which we assume to be outgassing) of $1.8{\times}10^{-14}$\,Torr\,l\,s$^{-1}$ (equiv. nitrogen) but gave no value for the glass surface area other than it being similar to Redhead et al's apparatus \cite{redhead1962ultrahigh} who used a 500\,ml bulb. Assuming the surface area of 300\,cm$^{2}$ we calculate an outgassing rate of $8{\times}10^{-17}$\,mbar\,l\,s$^{-1}$cm$^{-2}$. }
$10^{-16}$\,mbar\,l\,s$^{-1}$cm$^{-2}$.

\subsubsection*{Noble gases}
\begin{figure}
\includegraphics[width=0.45\textwidth]{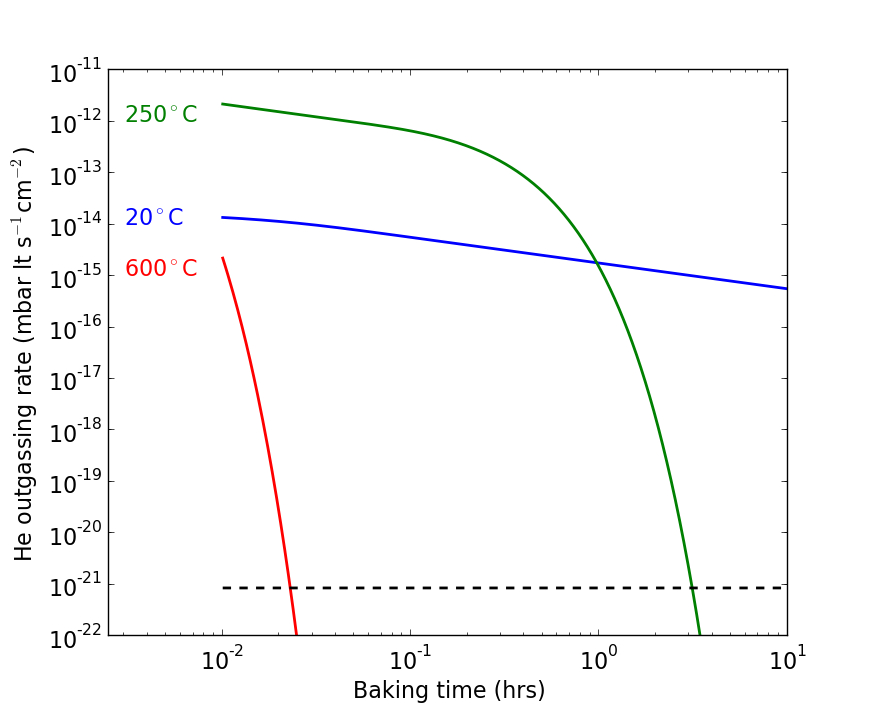}
\caption{\label{HeOG} The He outgassing rate from 1\,mm thick aluminosilicate glass slab at
various temperatures after storage in atmosphere (5\,ppm He content), calculated with Equation \ref{degaseq}. The black dashed line indicates
the target outgassing rate. }
\end{figure}

Noble gases cannot be removed once the MicroMOTs are sealed and so must be
\textit{completely} degassed from all components before bonding. As discussed in Section \ref{permsec}
negligible levels of noble gases, specifically helium, should be found in
silicon unless additional data corroborates a recent study \cite{sparks2013hermeticity}. Noble gases in glasses are expected to reach concentrations of 10\,ppb when exposed to atmosphere, which will outgas from the bulk diffusively. Figure \ref{HeOG} shows the results of degassing AS glass using Equation \ref{degaseq}, and we see that a thin slab can be completely degassed easily. This would scale proportionally to the area
when wafer level degassing is required. 

The last materials in our chips are metals such as gold for the atom chip
and those for the NEG. Noble gases do not permeate most metals and so one should be able to
ignore these materials, however the method of their deposition is important:
Sputtered thin films have been found to incorporate large quantities of argon, as
this process gas is used to remove the metal atoms from the sputter target. The argon
is then buried in the growing thin film and permeates to the surface along dislocations and pores, as
well as via self-diffusion. Where possible films should be deposited by
vacuum arc deposition \cite{sharma2008ti} or e-beam evaporation which do not require additional
gases. In situations where sputtering is unavoidable several modifications
can be made to reduce contamination by this gas source including \cite{amorosi2001study,window1993removing}: lowering
the sputter gas pressure, increasing the substrate temperature, reducing the deposition rate, post annealing, and finally changing to a heavier gas.
Using this last modification by replacing argon with krypton has been shown to reduce the gas incorporation by a factor of $10^{3}$ or
even lower, but moving to xenon shows little improvement \cite{window1993removing}. Measurements of TiZrV NEG films have detected krypton outgassing rates \cite{hsiung2009desorption} at the sensitivity limit of the
detector - down to $10^{-19}$\,mbar\,l\,s$^{-1}$cm$^{-2}$ after several activations- which is still too high for our device, but this could be further reduced with higher temperature anneals. 

\subsubsection*{Carbon monoxide}

Carbon monoxide is the second most significant gas at UHV. Due to its 
relatively large size (compared to helium or hydrogen) diffusion through 
the bulk will be negligible and so carbon monoxide is largely a surface outgassing species \cite{palosz2004residual}. 
Studies looking at
the residual gases in glass have shown that the carbon monoxide concentrations vary widely
\cite{redhead1962ultrahigh} and it is difficult
to obtain repeatable results. In addition, this also depends on glass type, processing history, and the effects of ionization cathodes. 
We do not know the desorption energy for carbon monoxide on glass, but due to the latter's low reactivity we expect carbon monoxide to 
only be physisorbed and so quickly degassed. Similar conclusions have been noted in photodesorption measurements \cite{lange1965photodesorption}. One study looking at the effect of breaking glass substrates in vacuum found that the carbon monoxide level remained constant, but carbon dioxide quickly decreased \cite{baptist1992carbon}. This was attributed to the reaction with residual hydrogen forming methane, which was seen to increase. 
A careful study \cite{bennett2004outgassing} avoiding the effects of
gauges measured carbon monoxide outgassing rates from stainless steel less than $6{\times}10^{-17}$\,mbar\,l\,s$^{-1}$cm$^{-2}$,
three orders of magnitude less than hydrogen.

Thermal desorption
studies of carbon monoxide released from silicon surfaces are few, but show that the
thermal desorption energy is below 0.5\,eV \cite{hu1997nonthermally} and so should be degassed easily. Photodesorption and plasma cleaning have also been shown to efficiently remove carbon and oxygen contamination from silicon surfaces.

\subsubsection*{Other gases}

Methane is also found in UHV environments and is thought to be produced from reactions on
the high temperature electrodes of ionization gauges, so should not be an issue in our gauge-less MicroMOTs.
It may also be formed from reactions between carbon monoxide and hydrogen during their diffusion on NEG or 
glass surfaces \cite{baptist1992carbon}, as mentioned before. TiZrV NEGs do not pump methane and so this gas 
should be completely evacuated before the MicroMOT is sealed, otherwise one must use reactive getters. Other organic species have been found in encapsulated MEMs-type devices which are likely due to insufficent cleaning or residual gases prior to sealing \cite{lindroos2009handbook, hasegawa2013effects}. 

Should anodic bonding be used to seal the chips, oxygen will be released along
the inner bonding edges where the voltage is high and no silicon exists to bond
with the non-bridging oxygen atoms \cite{knowles2006anodic, mack1997analysis, henmi1994vacuum}. Predicting the amount of oxygen released
is unreliable due to the lack of data and the effects of bonding parameters,
chip dimensions, and increased outgassing at raised temperatures. By analyzing the few studies on this subject \cite{mack1997analysis, henmi1994vacuum, rogers1992considerations} we estimate
$10^{13}$ to $10^{14}$ molecules per millimetre inner bonding circumference. For our
MicroMOT design this can lead to a monolayer coverage of oxygen on the NEGs
and result in saturation. It should be noted that oxygen penetrates the
NEG surface resulting in a capacity of about five monolayers \cite{chiggiato2006ti}.
Therefore it is important to maintain the chip at high temperatures after
bonding to absorb the oxidized NEG layers into the bulk. Once rubidium is
released into the chip it will quickly oxidize with any remaining oxygen forming
Rb$_2$O, which also reacts exothermically with water and hydrogen forming stable
hydroxides and hydrides which do not contaminate vacuum.

\subsection{Vacuum discussion}
\label{modelsec}

We have identified all the main sources of residual gases which could threaten
our sealed UHV environment. We have seen that helium permeation through glass
can be reduced to a negligible level with the use of aluminosilicates and could
further be improved with optical coatings such as graphene. Leaking through
bonds must be several orders of magnitude higher than has been measured, but can
be sufficiently improved by incorporating a `moat' within the
bonding seam. We also note that leakage can be further reduced by coating the
inner edges of bonding seams with NEG films and by applying a barrier coating on
the outer edges of the device. Several bonding techniques are available and we
highlight eutectic and direct bonding as the most suitable methods due to their
low outgassing and high hermeticity, with anodic bonding as a suitable
alternative if the oxygen released during bonding can be pumped away. Lowering the
temperatures of these bonding techniques
should be investigated as they can reduce the outgassing limitations by two or three orders
of magnitude \cite{shoji1998low, wei2003low, gerlach1999low, lee1993bonding, mescheder2002local, 
gosele1995self, takagi1998low, chuang2010study, choi2002analysis, dragoi2008wafer}.

\begin{table}
\caption{\label{lowout} Lowest and typical room temperature outgassing rates
for 2mm thick materials. The theoretical values (Th.) have been estimated using
Equation \ref{degaseq} and Figure \ref{h2bake} with a 10\,hr 250$^\circ$C vacuum bake for `typical
outgassing' and an additional 1\,hr 600$^\circ$C vacuum bake for `lowest outgassing'.}
\begin{tabular*}{0.45\textwidth}{@{\extracolsep{\fill} }lcc}
\toprule
Outgassing rate & Lowest & Typical \\
(mbar\,l\,s$^{-1}$cm$^{-2}$) &  &  \\
\midrule
Silicon\footnote{Theoretical values taken from Figure \ref{h2bake}.} &  $10^{-30}$ (Th.) & $10^{-24}$ 
(Th.)\\
Aluminosilicate & $10^{-17}$ & $10^{-16}$ (Th.\footnote{Using values from Table
\ref{diff}})\\
Pyrex  & $10^{-14}$ & $10^{-10}$\\
Stainless steel & $10^{-15}$\footnote{Thinner materials have lower outgassing
rates and Nemani{\v{c}} et al \cite{nemanivc1998outgassing} have demonstrated $10^{-17}$ with 150\,$\mu$m foil.} &
$10^{-12}$\\
\bottomrule
\end{tabular*}
\end{table}

The greatest hurdle we are left with is
to reduce outgassing. This can be tackled in two ways: 1) improve the pumping
rate and capacity of the getter films, and 2) reduce the outgassing rate by
extensive degassing procedures and careful choice of materials. 

It has been
shown that the pumping rate of NEGs is difficult to improve even with reactive
lithium getters, however the latter retains a constant pumping rate irrespective of
its history. NEGs are more straightforward to deposit, can be
used to coat surfaces to reduce outgassing and are stable in air. Reactive getters need to be deposited under vacuum and could result in unwanted coating on components in the
chamber. However, they have far higher capacities and can pump gases such as methane,
which NEGs cannot. Therefore we see a combination of NEGs and reactive getters as a good
compromise with the former activated during bonding and the latter activated
after bonding.

For the second method to tackle outgassing we
have seen in Table \ref{lowout} that at room temperature the materials we have chosen for
the device are more than adequate once degassed to achieve the
room temperature outgassing rate of $10^{-13}$\,mbar\,l\,s$^{-1}$cm$^{-2}$. When we consider the
outgassing during bonding at around 400$^\circ$C it can
increase by eight orders of magnitude and put stricter room temperature
rates of less than $10^{-16}$\,mbar\,l\,s$^{-1}$cm$^{-2}$ for bulk gettered gases
such as hydrogen, $10^{-18}$\,mbar\,l\,s$^{-1}$cm$^{-2}$ for surface gettered
gases such as carbon monoxide, and  less than $10^{-21}$\,mbar\,l\,s$^{-1}$cm$^{-2}$ for non-gettered
noble gases. We can see in Table \ref{lowout} that silicon outgassing is likely to be negligible compared to 
AS glass whose rate matches our target. When calculating this value we assumed a diffusion energy of 0.9\,eV. We can now be confident 
that AS glass will be the major source of hydrogen so if we use a more realistic 
value of 0.79\,eV (Table \ref{diff}) we lower our target to $10^{-15}$\,mbar\,l\,s$^{-1}$cm$^{-2}$, 
which is certainly achievable. Carbon monoxide outgassing is difficult to predict but we expect it to be far lower than hydrogen, as found in stainless steel \cite{bennett2004outgassing}.
We have seen that noble gases may be sufficiently removed from the
chamber material with realistic baking parameters and by using the separated wafer
fabrication method shown in Figure \ref{bond}, one can ensure the optimum baking regime
for each material.

\section {Prototype MicroMOT}
\label{protosec}
We are now at the stage where we can design a prototype MicroMOT and the
fabrication process which is currently under development and will be
characterized at a later date.  As discussed in Section \ref{MOTsec} we assume a
G-MOT type geometry using a 10\,mm diameter grating structure and a cavity volume of 
$15{\times}15{\times}3$\,mm$^3$ to avoid light scattering off sidewalls.
Around 90\% of the beam overlap volume is within 2.5\,mm of the grating surface,
and so this is a reasonable choice of height and is also feasible to fabricate
from silicon using deep reactive ion etching, wet etching,  machining, or
powder blasting \cite{wensink2000high}. We have consciously avoided designing the MicroMOT around a single application or manipulation technique, e.g. free-falling atom interferometers, or BECs on atom chips. 
This is because the MicroMOT will likely need to be adapted for the specific task, so we 
have chosen a simple generic design to demonstrate what is possible.
\vspace{3mm}

The chip is formed of four chambers: a large science chamber for
cooling and manipulating the atoms, an atom source chamber in which to hold the alkali dispensers,
an alkali getter/LIAD/peltier chamber, and a reactive getter chamber. The source
chamber is connected to the science chamber by a very thin channel ($1{\times}0.1{\times}0.1$\,mm) to restrict
the vapour flow, but experimental data will be needed to optimize these dimensions. 
The top 2\,mm thick capping wafer is anti-reflection coated and is anodically or direct bonded to the silicon `cavity wafer'. The thickness of the glass layer is determined by
several factors including the permeation rates, bondability, structural integrity and price. This wafer incorporates a 
moat within the bonding region on both sides to
reduce argon leakage. The reflector layer is coated by a thin alumina layer to
prevent alkali corrosion of the gold and is eutectically bonded to the cavity
wafer. We have not shown the quadrupole magnetic field coils as these are trivial to implement and may simply
be bonded, or deposited, onto the top and bottom external surfaces,
or could be approximated with a double loop on single surface \cite{jian2013double}.
\vspace{3mm}

\begin{figure}
\includegraphics[width=0.45\textwidth]{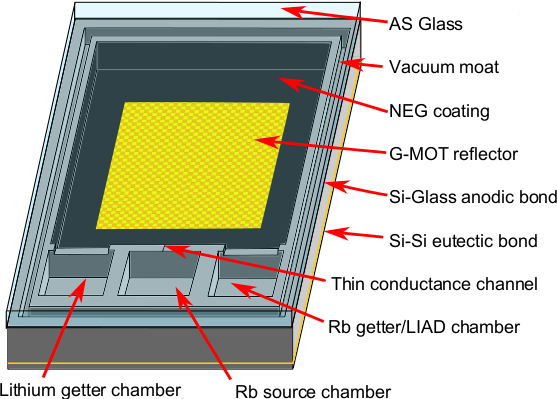}
\caption{\label{proto} Prototype design of a miniaturized magneto-optical trap
incorporating all the elements discussed in the text. Dimensions are
$20{\times}24{\times}5$\,mm$^3$.}
\end{figure}

Following extensive cleaning procedures, the
glass and silicon wafers are first bonded in vacuum so that oxygen released can be removed if anodic bonding is used. An NEG thin film is
sputtered on the internal sidewalls and onto some of the bottom layer (with the
reflector masked off) to provide the largest NEG area and reduce outgassing. The remaining reflector layer is then
eutectically bonded under UHV after high temperature baking to desorb
hydrogen, carbon monoxide and noble gases. 

Table \ref{protovac} provides a detailed summary of dimensions 
and an estimation on the vacuum properties of the device using values calculated in this study. We can see that the
lifetime of the device is determined by the argon leakage, but is nearly an order of magnitude greater than our target. 
The extremely low outgassing rate of AS glass results in negligible outgassing after bonding such that the base pressure 
is in the XHV regime for an essentially unlimited lifetime. Practically, the lifetime and base pressure will be determined 
by the atom source and the ability to pump away the vapour. The incorporation of high surface area materials will inevitably result in significantly increased outgassing, but porous silicon may be the better candidate due to the inherently low outgassing 
 rate of silicon and the ability to desorb hydrogen with UV light \cite{collins1992photoinduced}.       
\begin{table*}
\caption{\label{protovac} Expected vacuum properties of the prototype MicroMOT in atmosphere at $20^\circ$C.}
\begin{tabular*}{\textwidth}{@{\extracolsep{\fill} } lccc}
\toprule
Internal volume (cm$^{3}$) & 0.65 & including subchambers & \\
Surface area (cm$^{2}$) & 7.8 & 2.6 glass & \\
NEG\footnote{Surface capacity of $5{\times}10^{14}$\,molecules\,cm$^{-2}$ and bulk capacity of $7.5{\times}10^{16}$\,molecules\,cm$^{-2}$ (1\,$\mu$m film with $x_{H}=0.01$).} area (cm$^{2}$) & 3.3 & all sides and some of the reflector surface & \\
Glass thickness (cm) & 0.2 &&\\
Moat volume (cm$^{3}$) & 0.03 & $1{\times}0.5$ mm trench within each bonding seam&\\
Bonder base pressure (mbar) & $10^{-9}$ & equal H$_{2}$ and CO, negligible noble gases & \\
Bonding parameters & 1\,hr at 400$^\circ$C & typical eutectic bond & \\
\midrule
Gas source (mbar\,l\,s$^{-1}$)& Surface pumped (CO)& Bulk pumped (H$_2$) & Noble gas\\ 
\midrule
Permeation\footnote{The permeation rates are calculated using the breakthrough time from Equation \ref{permeation} substituted into Equation \ref{vaceq}.} & - &  $1.3{\times}10^{-23}$ &  $7.3{\times}10^{-23}$\\
Leakage\footnote{Bond leak rate (air) of $10^{-15}$\,mbar\,l\,s$^{-1}$ for 2 bonding seams (top wafer and bottom wafer). We have assumed the carbon 
monoxide leak is from atmospheric carbon dioxide.} & $1.9{\times}10^{-23}$ & $3.8{\times}10^{-24}$ & $2.3{\times}10^{-22}$ \\
Outgassing\footnote{As the carbon monoxide outgassing rate is unknown we have assumed a value which is ten times less 
than hydrogen as discussed in the text as a worst case scenario.} & $4.5{\times}10^{-17}$ & $2.6{\times}10^{-16}$ & $<10^{-24}$\\
\midrule
Base pressure\footnote{The base noble gas pressure will be equal to the residual level in the bonding chamber.} (mbar) & $10^{-18}$ &$10^{-15}$& - \\
Lifetime (days to reach $1.5{\times}10^{-10}$\,mbar) &$10^{6}$ & $10^{9}$&3200\\
\bottomrule
\end{tabular*}
\end{table*}

\section{Discussion and Conclusions}
\label{dissec}
We have shown that it is feasible to maintain UHV, and even XHV, environments for extended periods inside sealed chips using materials and methods borrowed from the semiconductor industry. 
However, it is necessary to highlight the assumptions we have made if these type of devices 
are to become a realistic technology. A great emphasis has been made on degassing the 
materials at temperatures up to 600$^\circ$C to ensure a sufficiently low outgassing rate once sealed.
 For the bulk materials discussed this is certainly possible, but more sophisticated devices are likely to
 have additional components, such as micro-Peltier coolers, detectors, field emission tips
 (for active pumping or ionization), or even light sources (such as UV LEDs for LIAD), which can be sensitive to extreme temperatures.
 Moreover, thin films, such as gold on silicon, can diffuse at moderate temperatures if 
additional barrier layers are not used \cite{lani2006gold}. In these situations one must use lower temperature degassing, 
such as UV desorption or plasma cleaning, and also develop lower temperature bonding methods \cite{shoji1998low, wei2003low, 
gerlach1999low, lee1993bonding, mescheder2002local, gosele1995self, takagi1998low, chuang2010study, choi2002analysis, dragoi2008wafer}. 

The very low leak rates we have predicted are possible by a combination of hermetic bonding and additional moat cavities. 
We have assumed that the materials to be joined are reliably homogeneous and intact (e.g. perfect
crystallinity in the case of silicon) 
but in reality fabrication processes may lead to defects which can result in additional leakage routes such as
microcracks, crystal plane dislocations, surface defects, or thin oxide films. Surface barrier films, NEG
coatings and stress-relief annealing can reduce these effects but these possible sources of leakage are still
worth bearing in mind.

We have used the large amount of data on hydrogen diffusion in silicon to predict that it has a very low outgassing rate. 
We find that this rate is consistently low whichever values we use from the literature, especially after a
high
temperature bake for several hours, and so we are confident in the estimate. However, as mentioned in Section \ref{outsec},
there is no absolutely reliable method to predict the outgassing rate from real materials and the simple 
diffusion-limited model is only useful to an order of magnitude at best, 
especially when considering the migration of reactive species such as hydrogen. This can be seen with studies looking
 at the outgassing of stainless steel, where the diffusion limited model produces reasonable estimate for low 
temperature bakes (below 300$^\circ$C), but generally fails to predict the effect of very high temperatures.
 This is usually attributed to the effect of surface oxides which are more stable during low temperature baking and
 act as diffusion barriers \cite{ishikawa1995importance}. Therefore we expect silicon to have a very low outgassing rate 
 but probably higher than the value stated in Table \ref{lowout}. Experimental studies focusing specifically
on outgassing are required.
We have used measured values for AS glass outgassing, but glasses are notorious for producing variable results \cite{robinson1968physical}
so it is necessary to perform additional outgassing studies on the specific glass one uses to ensure suitability. For the sake of bevity we have limited our discussions to silicon and glasses but there are likely to be many other suitable materials, most noteably ceramics which are equally suitable.
\vspace{3mm}

In conclusion, the aim of this study was to prove that Magneto-Optical Traps can be 
miniaturised and integrated into devices capable of leaving the laboratory. We
have shown that recent advances in microfabrication techniques and materials can
lead to sealed chambers with microlitre volumes that maintain UHV for at least 1000
days using only passive pumping elements. The MOT geometry can be miniaturized 
to use a single laser beam, patterned reflectors, and atom source.  Controlled by a 
number of methods including LIAD, integrated cold fingers, 
conductance channels and several pumping mechanisms. The main issues to maintain 
sealed UHV environments are the need for extremely 
low leakage bonds, low outgassing materials, and also negligible noble gas outgassing
from chamber walls and sputtered films. We hope that this work motivates the development 
of ultracold quantum technology which has a vast number of practical
applications and promises to be a fruitful technology in a number of fields.

We would like to acknowledge Tim Freegarde, David Smith, Erling Riis, Aidan
Arnold, Joe Cotter, Wolfgang Reinert, and Kai Bongs, for their helpful and
insightful advice and ideas in the development of this study. This work was
supported by the Royal Academy of Engineering, EPSRC, the Royal Society, and
DSTL.


%
%

%

\bibliography{UHVrefs}

\end{document}